\documentclass{aa}
\usepackage{graphicx}
\usepackage{txfonts}
\usepackage{soul}
\usepackage{ulem}
\begin{document}

   \title{Modelling of asymmetric nanojets in coronal loops}

   \subtitle{ }

   \author{P. Pagano\inst{\ref{inst1},\ref{inst2}} \and
           P. Antolin\inst{\ref{inst3}} \and
           A. Petralia\inst{\ref{inst2}}
          }

\authorrunning{Pagano, Antolin, Petralia}
\titlerunning{Asymmetric jets}

\institute{Dipartimento di Fisica \& Chimica, Universita\' di Palermo, Piazza del Parlamento 1, I-90134 Palermo, Italy \label{inst1}
\and
INAF-Osservatorio Astronomico di Palermo, Piazza del Parlamento 1, I-90134 Palermo, Italy \label{inst2}
\and
Department of Mathematics, Physics and Electrical Engineering, Northumbria University, Newcastle upon Tyne, NE1 8ST UK \label{inst3}
\\
\email{pp25@st-andrews.ac.uk}
}

   \date{ }
 
  \abstract
   {Observations of reconnection jets in the solar corona are emerging as a possible diagnostic to study highly elusive coronal heating. Such jets, and in particular those termed nanojets, can be observed in coronal loops and they have been linked to nanoflares. However, while models successfully describe the bilateral post-reconnection magnetic slingshot effect that leads to the jets, observations reveal that nanojets are unidirectional, or highly asymmetric, with only the jet travelling inward with respect to the coronal loop's curvature being clearly observed.}
   {The aim of this work is to address the role of the curvature of the coronal loop in the generation and evolution of asymmetric reconnection jets. }
   {In order to do so, we first use a simplified analytical model where we estimate the post-reconnection tension forces based on the local intersection angle between the pre-reconnection magnetic field lines and on their post-reconnection retracting length towards new equilibria. Second, we use a simplified numerical magnetohydrodynamic (MHD) model to study how two opposite propagating jets evolve in curved magnetic field lines.}
   {Through our analytical model we demonstrate that in the post-reconnection reorganised magnetic field, the inward directed magnetic tension is inherently stronger (up to 3 orders of magnitude) than the outward directed one and that, with a large enough retracting length, a regime exists where the outward directed tension disappears, leading to no outward jet at large, observable scales. Our MHD numerical model provides support for these results proving also that in the following time evolution the inward jets are consistently more energetic. The degree of asymmetry is also found to increase for small-angle reconnection and for more localised reconnection regions.}
   {This work shows that the curvature of the coronal loops can play a major role in the asymmetry of the reconnection jets and inward directed jets are more likely to occur and more energetic than the corresponding outward directed ones.}

   \keywords{}

   \maketitle

\section{Introduction}

The solar corona is a very dynamic and variable layer of the solar atmosphere, where strong magnetic fields continuously drive and shape the million degree coronal plasma.
Most of the plasma in the solar corona is embedded in curved magnetic structures, coronal loops \citep{2014LRSP...11....4R}, that connect at their footpoints with the solar chromosphere.
Coronal loops are known to be heated to millions of degrees by energy release processes that are impulsive in nature, and a product of the dissipation of magnetic energy. However, the temporal and spatial distribution of these events and the exact physical processes involved are still strongly debated \citep{Klimchuk_2015RSPTA.37340256K}.

One of the most scrutinised scenarios for coronal heating is commonly known as the nanoflare model. Due to the continuous shuffling from magnetoconvection, the magnetic field lines composing coronal loops are expected to be braided at sub-arcsecond resolution \citep{vanBallegooijen2011}. Parker envisioned that this process would eventually lead to the development of tangential discontinuities or tiny current sheets ubiquitously in the corona, where magnetic reconnection would occur and release tiny amounts of energy in the nanoflare range \citep{Parker1988}. If frequent enough, such nanoflares may account for the heating of coronal loops \citep{Hudson_1991SoPh..133..357H}. In the Parker model, magnetic reconnection is driven by the small misaligned transverse components of the field with respect to the dominant guide field, and it is therefore also known as component magnetic reconnection. The dissipated magnetic energy is turned into thermal and kinetic energies, as well as particle acceleration.

For decades, observations and models have focused in ways to isolate nanoflares by detecting either the in-situ  sudden surge in temperature that follows the heating, or the effect of it on the transition region via accelerated particles or thermal conduction. Small nanoflare-like intensity bursts have been detected in multi-wavelength observations in the upper transition region/low corona \citep[e.g.][]{Testa_2013ApJ...770L...1T, Testa_2014Sci...346B.315T, Tian_2014ApJ...790L..29T}, and high temperatures of $10^7~$K have been indirectly inferred from X-ray observations \citep{Ishikawa_nature_2017}, all interpreted as the result of coronal nanoflares. Yet, this has proved insufficient to establish a direct link to the heating mechanism due to the fact that nanoflare-like intensity bursts are not unique to magnetic reconnection, with wave-based heating mechanisms also resulting in such episodic heating \citep{Moriyasu_2004ApJ...601L.107M,Antolin_2008ApJ...688..669A}. 

Recently, \citet{AntolinNature} have shown that a reconnection-based nanoflare has an observable dynamic counterpart: the nanojet. Nanojets are confined (widths and lengths of of 500~km and $1,000-2,000$~km, respectively), shortlived (on the order of 15~s or less) and very fast ($100-200$~km~s$^{-1}$) plasma flows perpendicular to the coronal loop guiding magnetic field. A myriad nanojets were detected in an avalanche-like spatial and temporal progression, leading to the formation of a hot coronal loop. By conducting numerical simulations, nanojets were shown to be caused by the slingshot effect during reconnection, i.e. the perpendicular magnetic tension component that is rapidly generated in the aftermath of magnetic reconnection that accompanies the nanoflare, and that is often invoked for reconnection jets. One of the most peculiar characteristics of the nanojets is that most point radially inward with respect to the curvature of the loop, thereby being singular or unidirectional (asymmetric) with respect to the reconnection point, in contrast with the bi-directional nature (symmetric) usually expected in the standard reconnection scenario. It was stated, although not therein proven, that this was due to the curvature of the coronal structure, a statement that we hereby aim to prove.

Magnetic reconnection is a common phenomenon in the heliosphere and observational evidences of its occurrence have been inferred. Jets are often interpreted as a manifestation of magnetic reconnection. The plasma is ejected outwards from the reconnection region
\citep[e.g.][]{1984GMS....30....1A}, accelerated by magnetic tension to Alfv\'enic speeds, often reflecting the high-$\beta$ or low-$\beta$ conditions of the environment \citep{2005AIPC..784..153S}. A non-exhaustive list of examples include photospheric jets and associated Ellerman bombs \citep[e.g.][]{2013ApJ...779..125N}, chromospheric jets \citep{2007Sci...318.1591S, 2017A&A...605A..49C}, type II spicules \citep{2017Sci...356.1269M}, surges, coronal jets associated to Coronal Mass Ejections \citep{2020SoPh..295...27S}, reconnection outflows during flares \citep[e.g.][]{Takasao_2012ApJ...745L...6T} or in the magnetopause \citep{2020JGRA..12527296M}.
Such outflows can show asymmetries caused by the initial global magnetic topology \citep[e.g.][]{2005AIPC..784..153S}, or by the weakly ionised plasma \citep{2015ApJ...805..134M}, geometrical asymmetries in the initial X-point configuration \citep{2007PhPl...14j2114C}, or complexities in the initial magnetic configuration  \citep{2013ApJ...769L..21A}.

The nanojet case differs from the more commonly investigated reconnection jets we have listed above for three reasons.
First, in a nanojet, the magnetic reconnection is limited to the component perpendicular to the guide field and it is therefore a small angle reconnection in a configuration where no opposite polarities are interacting. Second, the largest observed dynamics in the system are transverse to the field, with the field-aligned component smaller by an order of magnitude. 
Third, in the reconnection jet description, we consider the reconnection between magnetic field lines that are curved, due to the loop structure, or twisting or braiding, instead of focusing on local straight field lines around the reconnection point. We here show that this can be the main reason behind the jet asymmetry.

Observationally, the asymmetry of the jets consists with them being unidirectional in the plasma displacement perpendicular to the loop structures. In our modelling, we adopt a more general operational definition of asymmetric jets, as we investigate the causes of the observational signatures. In particular, we consider reconnection jets asymmetric if their outflow velocities from one side and the other of the X-point are significantly different, as this inevitably leads to higher displacement and stronger compression of the background medium. In a regime where the density does not vary in time, valid proxies for the asymmetry of the outflow velocities are the ratio between the forces exerted at either jet or the ratio between their kinetic energies.

Besides explaining why the observed nanojets in \citet{AntolinNature} are asymmetric, we predict that most nanojets in coronal loops are inherently asymmetric features, either because of the curvature, twisting or braiding. In order to explain the dynamics of these reconnection nanojets we first introduce a simple and idealised geometrical model
where we illustrate how the loop curvature can become a key factor in determining the direction and symmetry of the jets.
Second, we use magnetohydrodynamics (MHD) simulations to corroborate our first analysis from a different point of view,
where the jets triggering is rather symmetric and the asymmetry between the jets can arise from the MHD evolution of the system.
In both studies we investigate the role of the reconnection angle and the size of the region involved in the jets in determining the asymmetry of the nanojets.

The structure of paper is as follows. In Sect.\ref{analytical} we introduce our geometrical model for the nanojets in coronal loops, in Sect.\ref{mhdmodelling} we introduce and discuss the MHD simulations and we finally discuss the results and draw conclusions in Sect.\ref{conclusions}.

\section{Analytical model}
\label{analytical}
The interpretation of jets being driven by magnetic reconnection boils down to small misalignments between  magnetic field lines.
In the case of nanojets by \citet{AntolinNature}
this is caused by braiding within coronal loops and in this context, the guiding magnetic field is the average field direction, coincident with what could be perceived as a main axis for the loop. Reconnection nanojets then occur when a pair of magnetic fields lines in equilibrium reconnect at small angles and
generate a new pair of magnetic field lines whose out-of-equilibrium configuration leads to a significant magnetic tension
in the direction perpendicular to the guiding magnetic field. Therefore, prior to the jets, i.e. in equilibrium, there's no plasma displacement perpendicular to the magnetic field, while we can neglect plasma motions along the magnetic field lines as they are irrelevant for this study. 

In order to show what is the effect of the initial curvature of the magnetic field lines on the dynamics of the resulting jets, 
we consider a simple 2D system that we assume in equilibrium. In this model, let us describe two magnetic field lines in a cartesian $(x,y)$ reference frame with
\begin{equation}
\label{loop1}
y_1\left(x\right)=\left(1+\epsilon\right)\sqrt{1-\left(\frac{x}{1-\epsilon}\right)^2}
\end{equation}
\begin{equation}
\label{loop2}
y_2\left(x\right)=\left(1-\epsilon\right)\sqrt{1-\left(\frac{x}{1+\epsilon}\right)^2}
\end{equation}
where $\epsilon$ is a parameter for $y_1$ and $y_2$ that determines
whether the two curves are distinct $(\epsilon>0)$ or they coincide $\epsilon=0$.
The polarities of the field lines are not specified here, but they are assumed to be the same in agreement with the loop braiding scenario. For the non trivial case ($\epsilon>0$), we find that the two curves have one point $P=\left(P_x,P_y\right)$ of intersection satisfying:
\begin{equation}
\label{intersectionx}
P_x=\sqrt{\frac{2\epsilon}{\frac{\left(1+\epsilon\right)}{\left(1-\epsilon\right)}-\frac{\left(1-\epsilon\right)}{\left(1+\epsilon\right)}}}
\end{equation}
\begin{equation}
\label{intersectiony}
P_y=y_1\left(P_x\right).
\end{equation}
In this geometrical construction the point of intersection between the two magnetic field lines is considered as the point where magnetic reconnection can be triggered.

\begin{figure}
\centering
\includegraphics[scale=0.45]{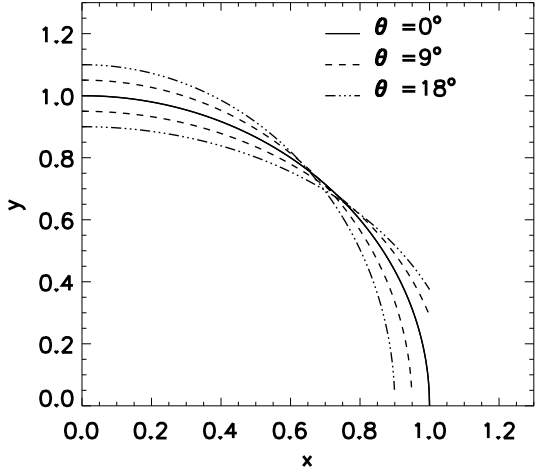}

\caption{Curves describing intersecting magnetic field lines at different angles from Eq.\ref{loop1} with $\epsilon=0$, $\epsilon=0.05$, and $\epsilon=0.1$.}
\label{fig1}
\end{figure}

The value of $\epsilon$ determines the angle $\theta$ between the two curves at the intersection $P$, so we choose here values of $\epsilon$ in the range $0\le\epsilon\le0.1$, leading to small misalignment angles.
Fig.~\ref{fig1} shows three pairs of such lines where $\epsilon$ is varied in order to have $\theta=0$ ($\epsilon=0$, i.e. two identical curves),
$\theta=9^{\circ}$ ($\epsilon=0.05$),
and $\theta=18^{\circ}$ ($\epsilon=0.1$).
We assume that this configuration is in equilibrium and thus that the magnetic tension at the point of intersection is somehow balanced by other forces. 
The magnetic tension exerted on the plasma depends on the local curvature radius of the magnetic field lines and, for this configuration, this could be calculated from the analytical expression of the magnetic field lines. However, we illustrate here an approximation we use that works also when such analytical expression does not exist (e.g. post-reconnection magnetic field lines).
From the intersection point $P$, we consider two points at distance $l$ along the $x$ axis and lying on $y_1\left(x\right)$
\begin{equation}
\label{epoints}
E_1^{\pm}=\left(P_x\pm l,y_1\left(P_x\pm l\right)\right)
\end{equation}
and similarly for $y_2\left(x\right)$.

Given 3 points on a $x-y$ plane with coordinates $(X1,Y1)$, $(X2,Y2)$, $(X3,Y3)$, the equation of the circle passing through these points can be written in terms of the following determinant
\begin{equation}
\label{determinantcircle}
\begin{vmatrix}
x^2+y^2 & x & y & 1 \\ 
X1^2+Y1^2 & X1 & Y1 & 1 \\ 
X2^2+Y2^2 & X2 & Y2 & 1 \\ 
X3^2+Y3^2 & X3 & Y3 & 1  \notag
\end{vmatrix}
=Ax^2+Ay^2+Bx+Cy+D=0
\end{equation}
where $A$, $B$, $C$, and $D$ are expressions of $X1$, $Y1$, $X2$, $Y2$, $X3$, and $Y3$ and can be derived expanding the determinant. The centre of the circle is located in 
$\left(-B/2A,-C/2A\right)$ and the radius is given by
\begin{equation}
\label{radiusABCD}
r=\sqrt{\frac{B^2+C^2-4AD}{4A^2}}.
\end{equation}

Using this analysis, we then find the equation of the circle $C_1$ passing through $E_1^{-}$, $P$, and  $E_1^{+}$
and the equation of the circle $C_2$ passing through $E_2^{-}$, $P$, and  $E_2^{+}$ and we use the radii of these circles as the curvature radius of the magnetic field lines in $P$.

We apply this analysis varying $l$ in the range $2^{-3}\le l\le0.2$.
We find that in our range of $\theta$ and $l$ the curvature radius ranges between $r=0.99$ and $r=1.13$.
We picked these ranges for $\theta$ and $l$, as we focus on small angle reconnection and relatively local effects. The maximum value of $l$ already corresponds to 20\% of the loop radius and higher values of the angle $\theta$ between the magnetic field lines is not significantly affected by the curvature radius.
In this method, as the magnetic field lines equations are not a circle, the parameter $l$ plays a role in the determination of the approximated curvature radius.
Fig.~\ref{fig2}a and Fig.~\ref{fig2}b illustrate this configuration for two different values of $\theta$.
Moreover, the centres of both circles lie internally with respect to both magnetic field lines.
As such a configuration is assumed in equilibrium and the magnetic tension depends on the curvature radius,
we associate curvature radius in this range with an equilibrium configuration.

\subsection{Post-reconnection magnetic field representation}

If we now focus on the effect of a magnetic reconnection event on the magnetic configuration at the point P,
we expect the magnetic field lines to become tangent at that point, instead of intersecting.
In particular, a new pair of curves is formed. The first one is composed partly by $y_1$ and partly by $y_2$, where we
take the two segments of either curves 
external with respect to the intersection point P.
The other field line is composed by the remaining parts of $y_1$ and $y_2$, so that this is always internal with respect to $P$. By construction, the curvature radius of these new curves is not analytically defined at $P$ since the derivatives are not continuous in this point where the analytical expression switches from Eq.\ref{loop1} to Eq.\ref{loop2} or vice versa. After the reconnection, the magnetic tension will tend to eliminate the said cusp and bring the new field lines to a new configuration.

Using the construction we have previously introduced,
the configuration switches from the pair of lines ($E_1^{-}$, P, $E_1^{+}$),  ($E_2^{-}$, P, $E_2^{+}$),
to the pair of lines ($E_1^{-}$, P, $E_2^{+}$),  ($E_2^{-}$, P, $E_1^{+}$).
When this occurs, in order to estimate the magnetic tension we compute the curvature radius of the 
circle $C_i$ defined by the points ($E_2^{-}$, P, $E_1^{+}$)
and the circle $C_e$ defined by the points ($E_1^{-}$, P, $E_2^{+}$), which again depends on the value of $l$.

\begin{figure*}
\centering
\includegraphics[scale=0.38]{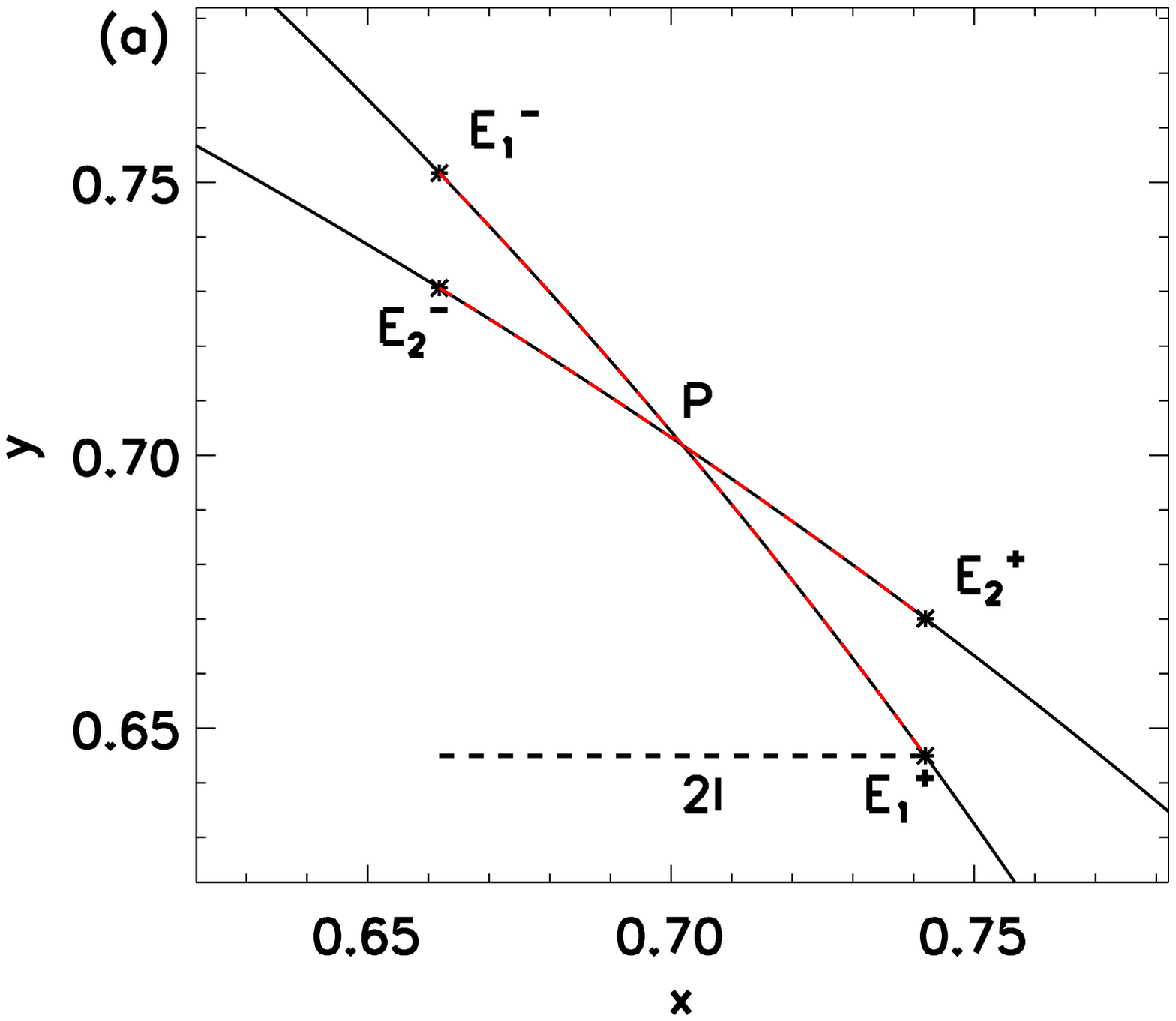}
\includegraphics[scale=0.38]{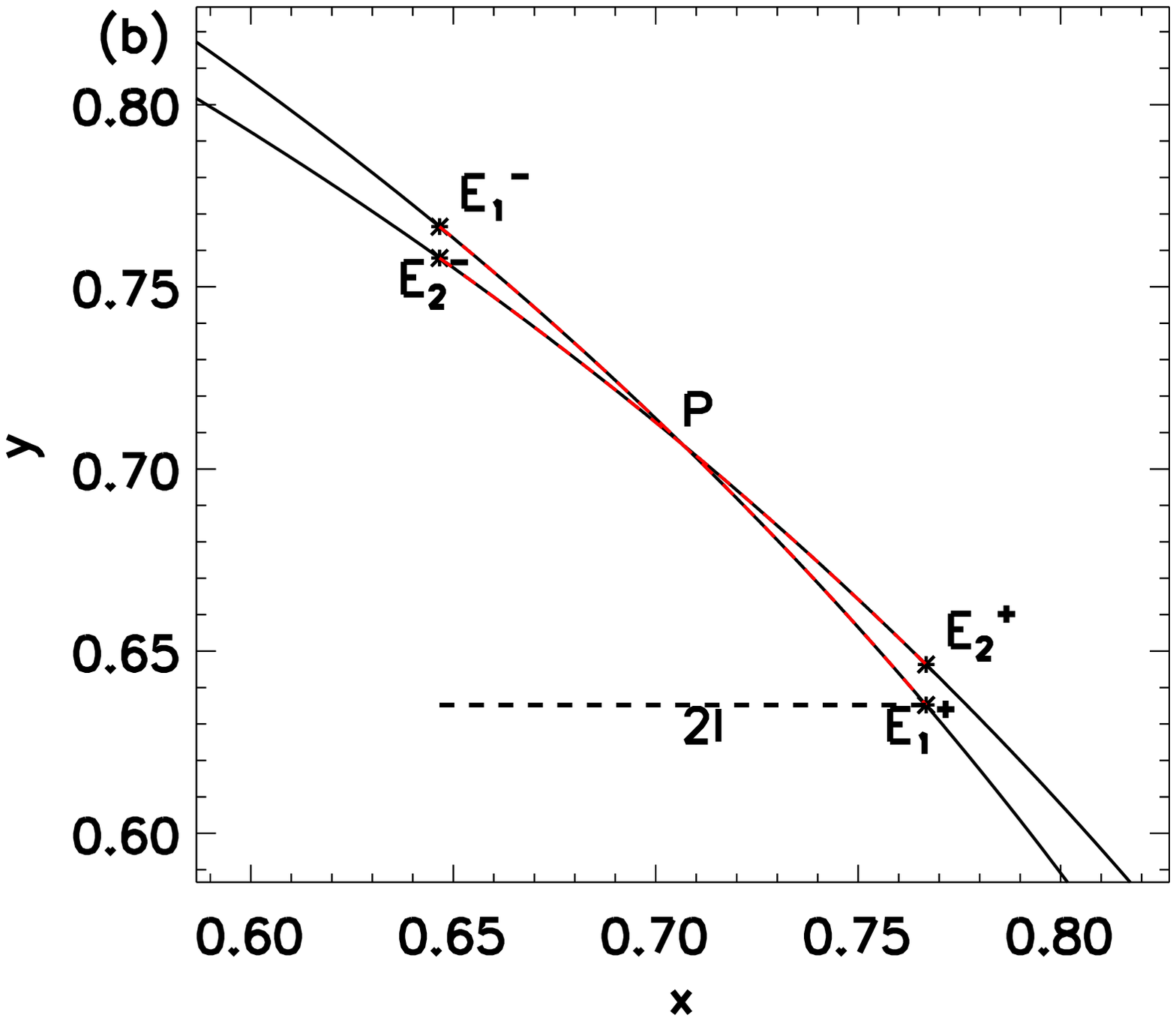}

\includegraphics[scale=0.38]{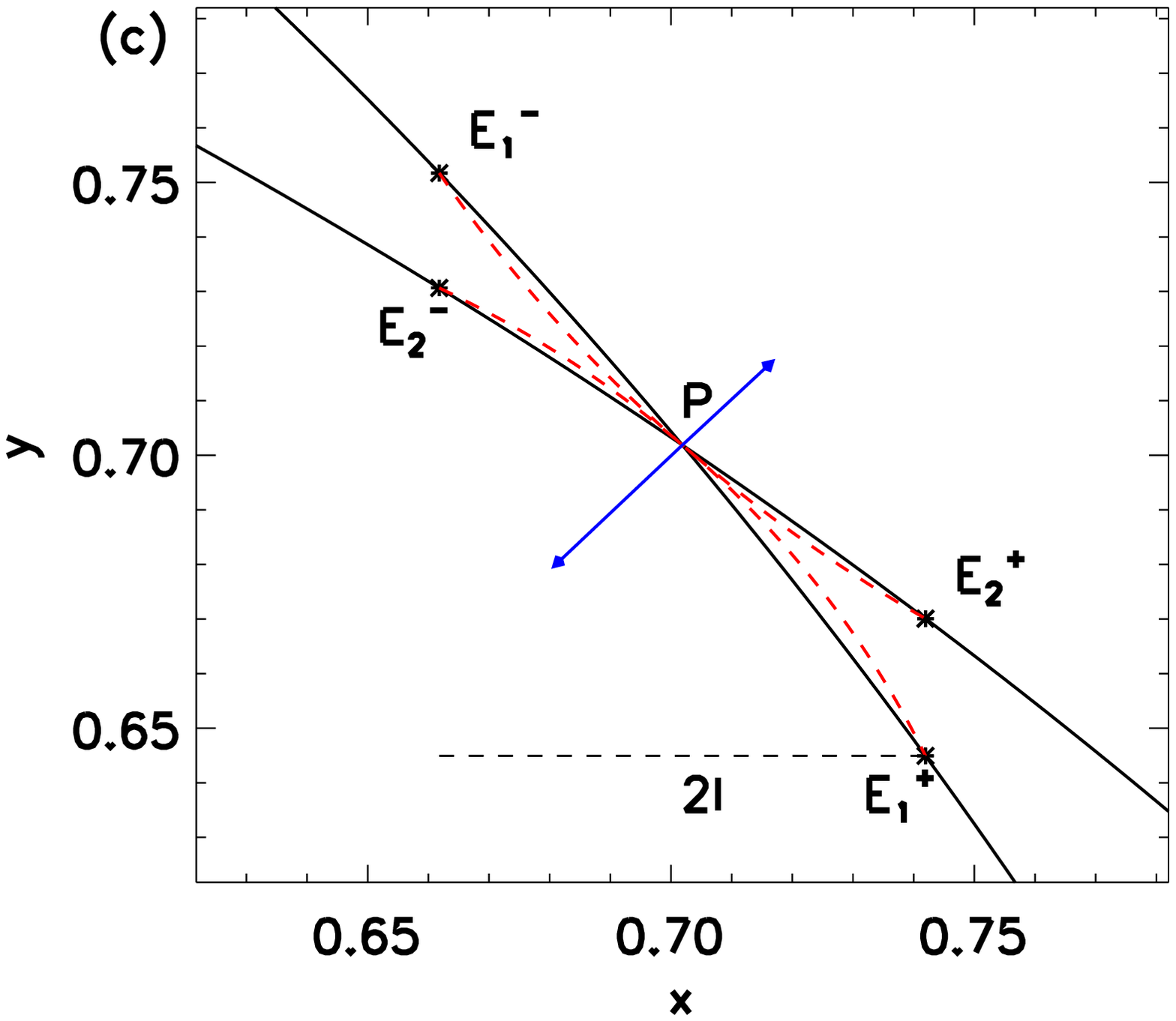}
\includegraphics[scale=0.38]{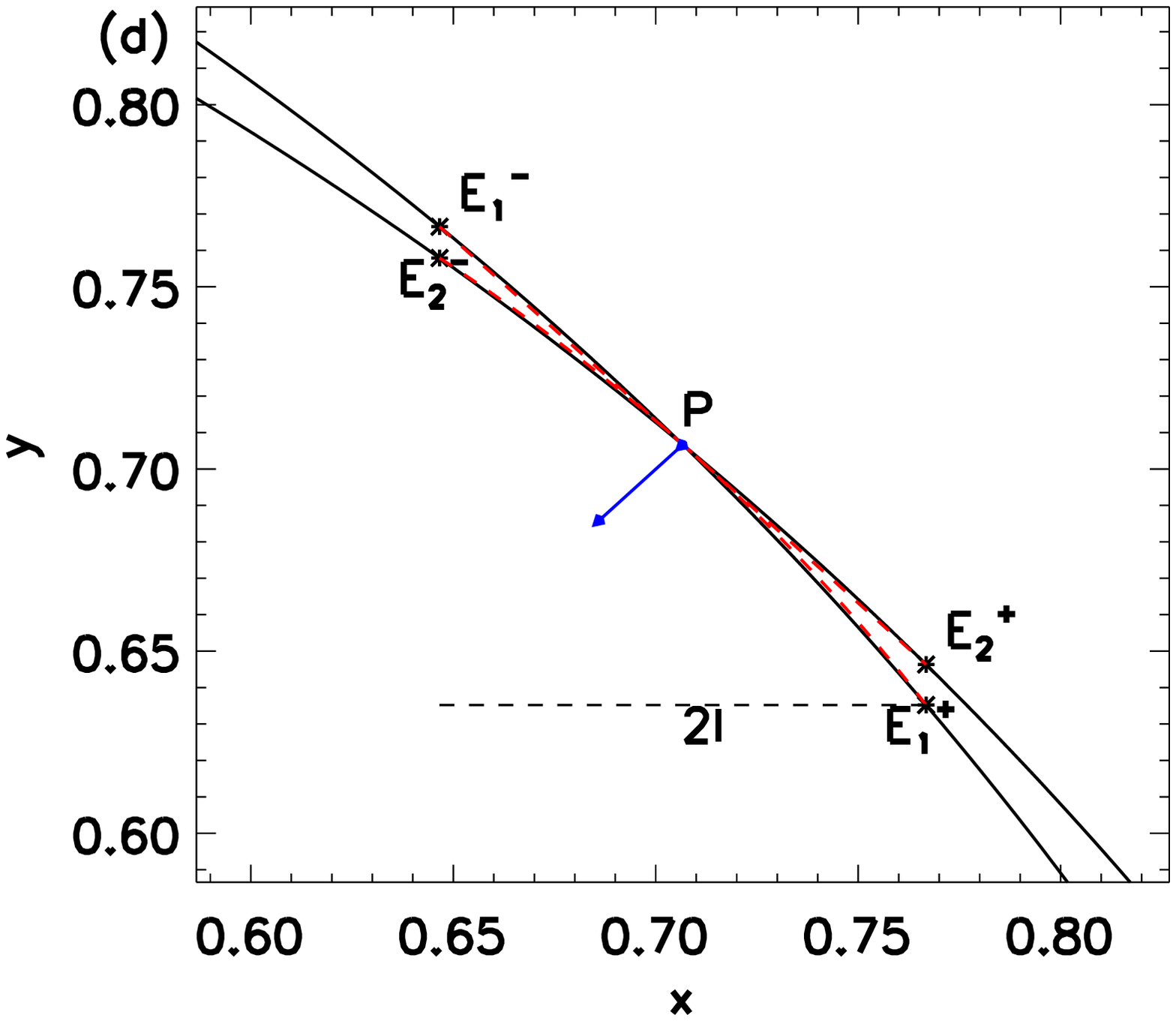}
\caption{Zoom in of the profiles of intersecting magnetic field lines at $\theta=13^{\circ}$ (panels a and c) and at $\theta=3.7^{\circ}$ (b and d).
Panels (a) and (b) show the configuration before the magnetic reconnection where the connectivity (red dashed arcs) connect
($E_1^{-}$, P, $E_1^{+}$) and ($E_2^{-}$, P, $E_2^{+}$).
Panels (c) and (d) show the configuration after the magnetic reconnection where the connectivity changed (red dashed arcs) and the connected points are ($E_2^{-}$, P, $E_1^{+}$) and ($E_1^{-}$, P, $E_2^{+}$). In panels (c) and (d) the length of the blue arrows is proportional to the inverse of the curvature radius of the red dashed arcs (defined by the circles passing through those points) and their directions point towards their centres of curvature.}
\label{fig2}
\end{figure*}

Fig.~\ref{fig2}c and Fig.~\ref{fig2}d illustrate this geometrical construction for the post magnetic reconnection configuration.
We find that two different scenarios are possible depending on the values of $\theta$ and $l$.
In the first scenario (Fig.~\ref{fig2}c) the centres of $C_i$ and $C_e$ are located in two different regions with respect
to the magnetic field lines. The centre of $C_i$ is the interior region, while the centre of $C_e$ is located externally with respect to
the magnetic field lines. Because of the curvature, for all values of $\theta$ and $l$ the radius $r_i$ of $C_i$
is smaller than the radius $r_e$ of $C_e$.
The magnetic tension intensity is inversely proportional to the local curvature radius and therefore, in this scenario, the inwardly directed magnetic tension is always stronger than the externally directed one. This is represented by the lengths of the blue arrows in Fig.~\ref{fig2}c, which are inversely proportional to the curvature radius. The direction of the blue arrows shows the direction of the resulting magnetic tension exerted by both magnetic field lines after reconnection. While this configuration would still allow for a bi-directional jet, the force generating is not symmetric.
In the second scenario (Fig.~\ref{fig2}d), the centres of both $C_i$ and $C_e$ are located in the interior region and no external magnetic tension is applied in this case.
In other terms, this configuration illustrates when both $E_1^{-}$ and $E_2^{+}$ lie in the interior part of the loops with respect to $P$, because of the loops curvature.
This happens when $l$ is large enough for the local curvature around $P$ to become negligible
for the circle $C_e$ and the overall curvature of the magnetic field lines system becomes dominant.

\begin{figure}
\centering
\includegraphics[scale=0.45]{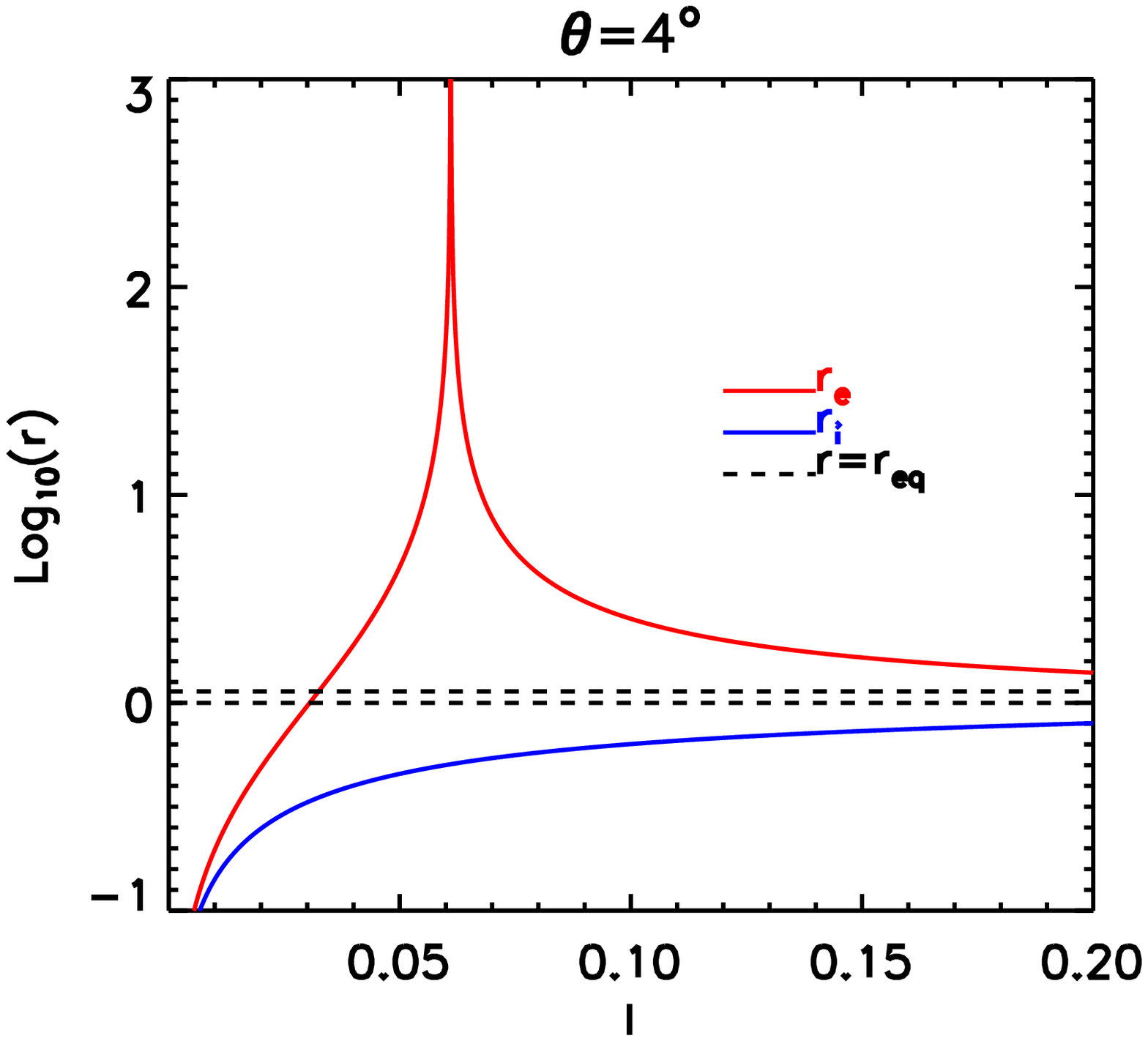}

\caption{Value of the curvature radius $r_i$ (blue line) and $r_e$ (red line) of the arcs in the post-reconnection field lines as a function of the length $l$.
The black dashed lines represent the minimum and maximum of the curvature radius of the
pre-reconnection magnetic field lines.}
\label{fig3}
\end{figure}

Fig.~\ref{fig3} shows the value of $r_i$ and $r_e$ in logarithm scale varying $l$ when $\theta=4^{o}$.
The dashed black lines show the limits of $r=0.99$ and $r=1.13$ that is the curvature radius range associated to the equilibrium configuration.
We find that for small values of $l$, $r_i$ is more than an order of magnitude smaller than the equilibrium radius
and only at large values of $l$ it approaches the equilibrium value.
In contrast, $r_e$ is initially very close to $r_i$, however it rapidly increases to values more than three orders of magnitude larger,
where it reaches a cusp maximum. This corresponds to when the three points ($E_1^{-}$, P, $E_2^{+}$) are in a straight line and we have no magnetic tension.
Past this maximum the convexity of the circle defined by ($E_1^{-}$, P, $E_2^{+}$)
changes and the exerted magnetic tension flips inward and $r_e$ decreases tending closer to equilibrium values.

\begin{figure}
\centering
\includegraphics[scale=0.40]{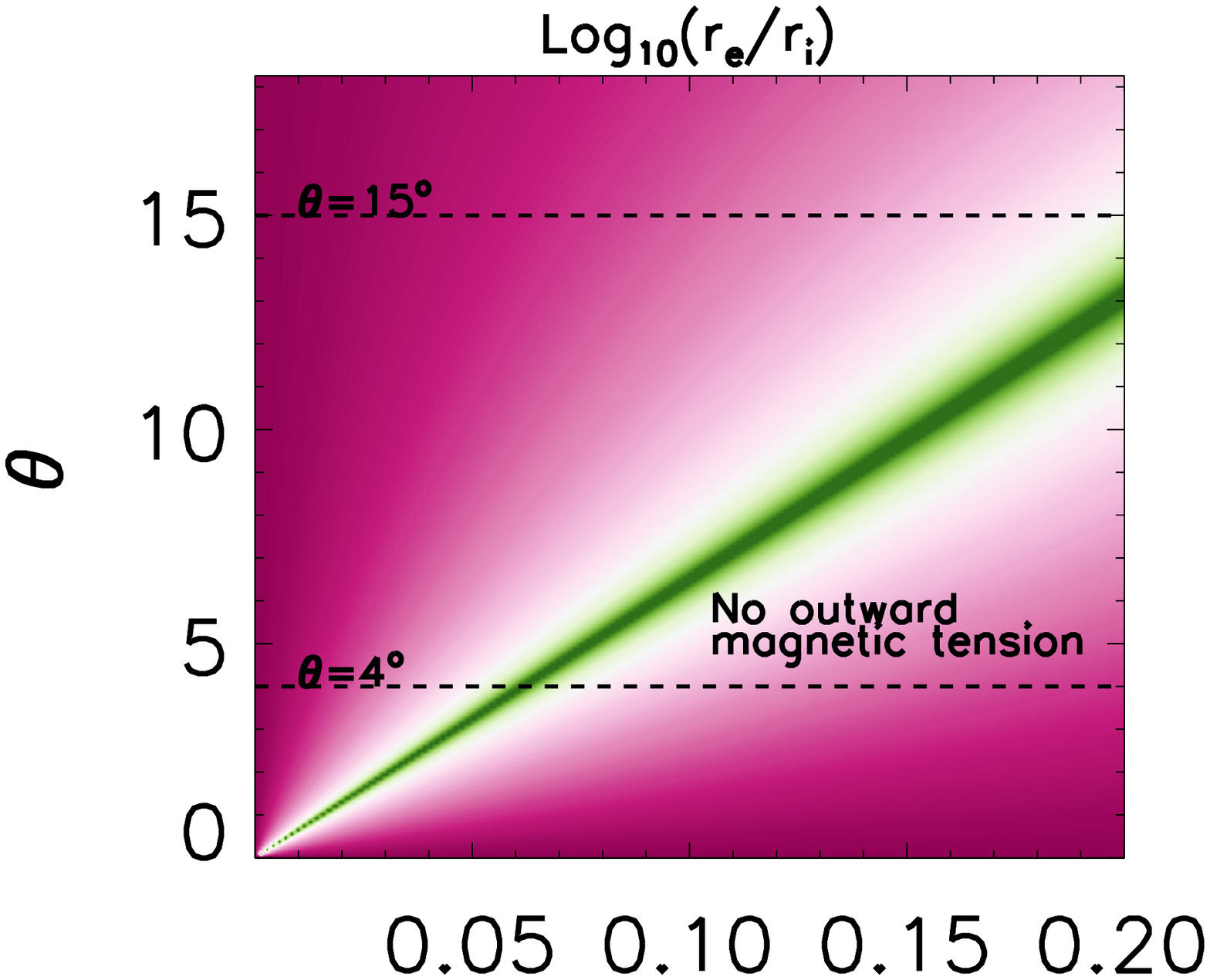}
\includegraphics[scale=0.40]{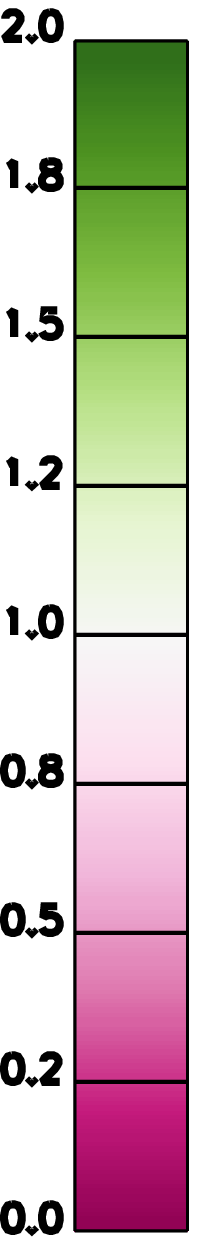}

\caption{Map of the Log of the ratio between $r_e$ and $r_i$,
as a function of the length $l$, and the angle $\theta$. The dashed lines indicate the cuts where $\theta=4^{\circ}$ and $\theta=15^{\circ}$.}
\label{fig4}
\end{figure}

Fig.~\ref{fig4} shows a colour map
of the ratio $r_e/r_i$ to illustrate the $\theta$ and $l$ dependence of the ratio between the inward directed and outward directed magnetic tension.
We find that where $C_e$ changes convexity the ratio $r_e/r_i$ reaches a maximum and this happens at larger $l$ the larger is $\theta$.
Hence, there is a very well defined region in the $\theta$-$l$ space that splits the diagram in two parts.
On the right hand side with respect to this region no outward magnetic tension is allowed and only an inward one can be considered.
On the left hand side, the magnetic tension that generates the inward jet is systematically larger than that generating the outward one. Only at very small values of $l$ the two forces are comparable.
However, for any $\theta$, as $l$ increases the inward magnetic tension becomes increasingly larger.

\subsection{Estimate of the time evolution}
\label{timeevolanalytical}

We now analyse the implication of these regimes for the time evolution of asymmetric jets.
In the following analysis of this simplified system,
we assume that the magnetic reconnection starts at the smallest spatial scale. Magnetic field lines start re-configuring by straightening up near the reconnection point, which changes the curvature radius (i.e. the magnetic tension), aiming for a new equilibrium.
This process therefore evolves encompassing larger regions and we define the distance involved in the reconfiguration by the retracting length.

\begin{figure*}
\centering
\includegraphics[scale=0.30]{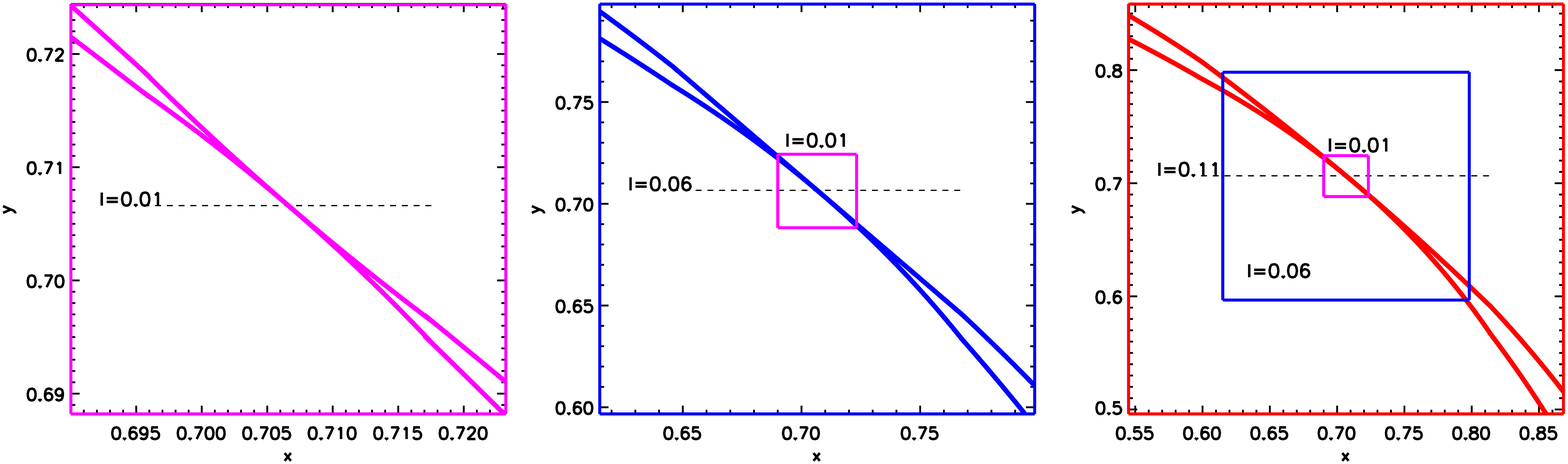}

\caption{Configuration of the reconnected field lines when $\theta=4^{o}$ at different retracting lengths $l$. In each panel we zoom on a different box around the point $P$ and we identify with coloured boxes smaller boxes. The red panel is the larger box, then blue, and the magenta is the smallest.}
\label{figrec}
\end{figure*}

In our analysis the retracting length is represented by the parameter $l$. This allows us to estimate a time evolution based on the assumption that the retracting length is linearly dependent with time. This can be justified if we assume that the geometry of the field lines (and thus the curvature) does not vary greatly around the reconnection region.
Fig.~\ref{figrec} can thus be seen as a time evolution of the post-reconnection evolution of the magnetic field lines.
When $l=0.01$ the global curvature of the magnetic field lines cannot be locally appreciated and the field lines form an x-shape, but the global curvature becomes more locally relevant as we move to larger retracting lengths at $l=0.06$ and $l=0.11$. The panel in the centre ($l=0.06$) corresponds to the configuration when the external magnetic field line changes convexity, so that we have a bidirectional asymmetric magnetic tension for smaller values of $l=0.06$, and unidirectional internally directed magnetic tension from both field lines for larger values of $l=0.06$.

\begin{figure}
\centering
\includegraphics[scale=0.45]{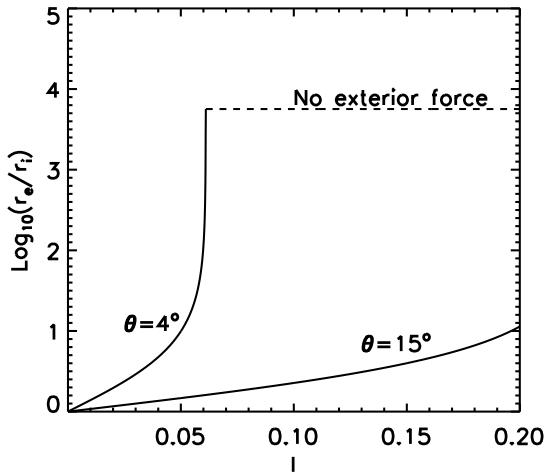}

\caption{Log of the ratio between $r_e$ and $r_i$
for two different angles $\theta$.}
\label{fig5}
\end{figure}

Fig.~\ref{fig5} shows the logarithm of the ratio between the inward directed magnetic tension and the outward directed one
for two different tilt angles, a small angle case ($\theta=4^{o}$) and a large angle case ($\theta=15^{o}$).
We find that in the small angle case, the ratio quickly reaches values of $10^3$
and the outward magnetic tension disappears shortly after,
whereas for larger tilt angle the increase is slower and in this range of $l$ the outward magnetic tension does not disappear.

In this model we assume the acceleration to be proportional to the force and to be positive away from the reconnection point and negative towards it.
Therefore, the force $f_i=1/r_i$, i.e. the magnetic tension exerted by the post-reconnection field line ($E_2^{-}$, P, $E_1^{+}$) is always positive, whereas $f_e=1/r_e$, the magnetic tension exerted by the magnetic field line ($E_1^{-}$, P, $E_2^{+}$), can be either negative or positive.
Additionally, we assume the force $f_i$ to be null when the curvature radius is larger than $r_i=0.99$ that is the threshold for equilibrium identified in Fig.~\ref{fig3} and the force $f_e$ to be null when it falls in the regime where no external magnetic tension is allowed.

We then assume that the length $l$ linearly increases with respect to time $t$ and 
we integrate the equation of motion to find the ratio between the velocities ($v_i/v_e$) and the lengths ($S_i/S_e$) of the jets as a function of time using a Runge-Kutta scheme.
Thus, we have for the velocities
\begin{equation}
\frac{v_i}{v_e}=\frac{\int_0^t f_i dt^{\prime}}{\int_0^t f_e dt^{\prime}}
\end{equation}
and for the lengths of the jets
\begin{equation}
\frac{S_i}{S_e}=\frac{\int_0^t v_i dt^{\prime}}{\int_0^t v_e dt^{\prime}}
\end{equation}
We solve the equation of motion between $t=0$, when 
the jets speeds and lengths are 0 and the normalised time $t$, corresponding to our maximum values of $l$.

\begin{figure}
\centering
\includegraphics[scale=0.45]{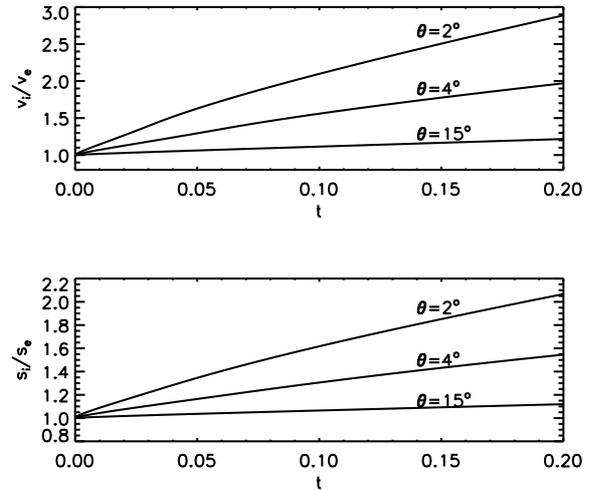}

\caption{Time evolution of the ratio between the speed of the internal and external jets (top panel) and their displacement (lower panel), for three different values of the angle $\theta$.}
\label{fig6}
\end{figure}

Fig.~\ref{fig6} shows the ratio between the inward and outward velocities and displacement as a function
of time for the small and large title angles.
We find that for small tilt angles the inward jets accelerates to speeds three times faster than the outward jet,
whereas the displacement is up to two times larger.
The asymmetry of these jets is consistent at all times and for all the tilt angles $\theta$ considered. The substantial asymmetry we measured in the tension force is not reflected in this time evolution because that is limited to a small region of the $\theta$-$l$ space. At the same time, we find the asymmetry of the time evolution of the jets to increase significantly for smaller tilt angles, indicating a non-linear evolution.

\begin{figure}
\centering
\includegraphics[scale=0.45]{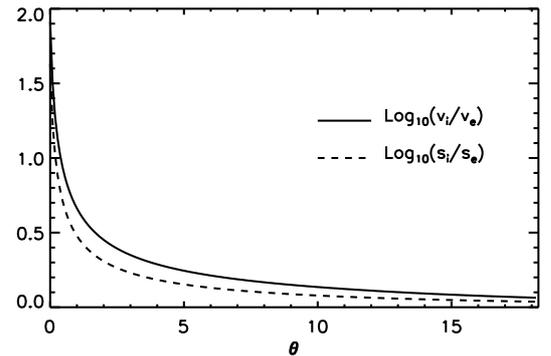}

\caption{Ratio of the Log of the final ratio of the speed (continuous curve) and displacement (dashed curve)
between the two jets as a function of the angle $\theta$.}
\label{fig7}
\end{figure}
Fig.~\ref{fig7} shows the final asymmetry in velocity and displacement in our time evolution for various values of the tilt angle $\theta$.
We find that the asymmetry is small when $\theta>10$,
but it becomes two orders of magnitude larger as we move to the small angle reconnection regime.
Of course, as the tilt angle approaches 0, we approach the limit where no reconnection jet should take place, but this results shows that it is possible to find a regime where reconnection and jets are taking place and the curvature of the loops leads to an asymmetric evolution.
Although this is a very simple model it provides a sense on how the initial misalignment of the curved magnetic field lines significantly affects the asymmetry in the dynamics and lengths of the resulting jets.

It should be noted also that the analysis performed here can equally apply to braided field lines, for which, locally, different curvatures exist. The small angle reconnection regime would be particularly applicable to braided field lines. The directions and speeds of nanojets are therefore likely linked to braiding as well as global loop curvature.

\section{MHD modelling}
\label{mhdmodelling}

In order to investigate the evolution of reconnection jets from curved magnetic field lines from a different perspective, we devise an MHD model where two magnetic flux systems show interlaced magnetic field lines in 3D configuration that has some commonalities with what described in Sect.\ref{analytical}.
The aim of these numerical experiments is to verify that the general relations we derived in Sec.\ref{analytical} between the asymmetry of the reconnection jets and the geometry of the magnetic configuration hold from an MHD perspective as well,
even starting from a mostly symmetric initial configuration.
We are not attempting at modelling solar corona asymmetric reconnection nanojets, as we do not describe the coronal plasma quantities or their evolution.

To develop this model we use the PLUTO code \citep{Mignone_2012ApJS..198....7M},
where the following ideal MHD equations are solved numerically.
\begin{equation}
\label{mass}
\displaystyle{\frac{\partial\rho}{\partial t}+\vec{\nabla}\cdot(\rho\vec{v})=0},
\end{equation}
\begin{equation}
\label{momentum}
\displaystyle{\frac{\partial\rho\vec{v}}{\partial t}+\nabla\cdot\left(\rho\vec{v}\vec{v}-\vec{B}\vec{B}+p_t\mathbf{I}\right)=0},
\end{equation}

\begin{equation}
\label{induction}
\displaystyle{\frac{\partial\vec{B}}{\partial t}+\nabla\cdot\left(\vec{v}\vec{B}-\vec{B}\vec{v}\right)=0},
\end{equation}

\begin{equation}
\label{energy}
\displaystyle{\frac{\partial E}{\partial t}+\nabla\cdot\left(\left( E+p_t\right)\vec{v}-\vec{B}\left(\vec{v}\cdot\vec{B}\right)\right)=0},
\end{equation}

where $t$ is time, $\vec{v}$ the velocity,
$p_t$ the total pressure, i.e. the sum of gas pressure $p$ and magnetic pressure $p_m$,
$\vec{B}$ is the magnetic field,
$J=\frac{c}{4\pi}\nabla\times\vec{B}$ is the current density,
$c$ is the speed of light,
and $\mathbf{I}$ is the identity tensor.
The total energy density $E$ is given by
\begin{equation}
\label{enercouple}
\displaystyle{E=\frac{p}{(\gamma-1)}+\frac{1}{2}{\rho}\vec{v}^2+\frac{\vec{B}^2}{2}},
\end{equation}
where $\gamma=5/3$ denotes the ratio of specific heats.

\subsection{Initial conditions and numerical setup}
In our model, the initial configuration consists of two shifted arcade systems defined in a cartesian
reference frame $(x,y,z)$ that 
extends from $x=-50$~Mm to $x=50$~Mm in the $x-$direction,
from $y=2.5$~Mm to $y=52.50$~Mm in the $y-$direction,
and from $z=-0.78125$~Mm to $y=+0.78125$~Mm in the $z-$direction.
The arcade systems develop on the $x-y$ plane and are defined by the magnetic field components
\begin{equation}
\label{magx}
B_x=+\pi \cos{\left(x\pm x_c\right)\pi} e^{-y\pi}
\end{equation}
\begin{equation}
\label{magy}
B_y=-\pi \sin{\left(x \pm x_c\right)\pi} e^{-y\pi}
\end{equation}
where $\pm x_c$ is a model parameter that is taken with its positive value for $z\ge0$ and its negative value for $z<0$.
In this way two identical configurations shifted by $2x_c$ along the x-axis are coexisting in the domain and one switches
into the other across the plane $z=0$.
Moreover, the angle between the magnetic field vector defined in $z<0$ and the one defined in $z>0$ is uniform.
The general direction of the magnetic field lines does not change, so that the two flux systems show only a small misalignment between each other.

In this model we adopt a configuration where the two flux systems are locally magnetically connected, so that the magnetic reconnection has already occurred around the point ($x_J=0$,$y_J=25$,0).
In order to describe this post-reconnection configuration we add a $z$ component of the magnetic field
\begin{equation}
\label{addbz}
B_z=B_{z0}e^{-\left(\frac{x-x_J}{L_{\parallel}}\right)^2}
e^{-\left(\frac{y-y_J}{L_{\perp}}\right)^2}\left(\frac{y-y_C}{L_{\perp}}\right)
\end{equation}
where $B_{z0}=1.1$~$G$, $L_{\perp}=0.25$~Mm, and $L_{\parallel}$ is a parameter that controls the extension in the $x$ direction of the region in which this connecting component is present. 

The $x$ and $y$ components of the magnetic field are force free by construction, therefore adding this z-component of the magnetic field generates unbalanced magnetic pressure gradient 
and magnetic tension.
In particular, the inward magnetic tension (towards the origin of the axes) is slightly (2\%) more intense than the outward magnetic tension (away from the origin).
Such construction is clearly different from the analytical one, where we estimated orders of magnitude difference between the inward and outward magnetic tension.
With this numerical experiment, we aim at showing that also the MHD evolution that follows the triggering of the jets adds asymmetry to the jets evolution.

The density in our initial conditions is defined as
\begin{equation}
\rho=2\rho_{D}+\frac{\left(0.5\rho_0-\rho_D\right)}{2}
\left[\left(1+\tanh{\left(y-5\right)}\right)+
\left(1+\tanh{\left(50-y\right)}\right)\right]
\end{equation}
where $\rho_0=4.2$~g~cm$^{-3}$ and $\rho_D=10^3\rho_0$.
With such a density distribution we have a density value of $\rho_D$ in two regions of $5$~Mm in width near the two $y-$boundaries and $\rho_0$ elsewhere. The purpose of these high density regions is
to slow down the propagation of perturbations from the $y-$boundaries such as to avoid artificial boundary effects.
Finally, the thermal pressure distribution is constructed to have a uniform total pressure $p_T=p+\vec{B}^2/8\pi=0.42$~erg~cm$^{-3}$, corresponding to a value of $\beta=p/(B^2/8\pi)=0.47$ around the location ($0$,$y_J$).
In this configuration, only the magnetic tension introduced by $B_z$ remains as unbalanced force that can drive the dynamics of the numerical experiment.
Finally, the temperature is derived from the pressure and density from the ideal equation of state.

\begin{figure*}
\centering
\includegraphics[scale=0.35]{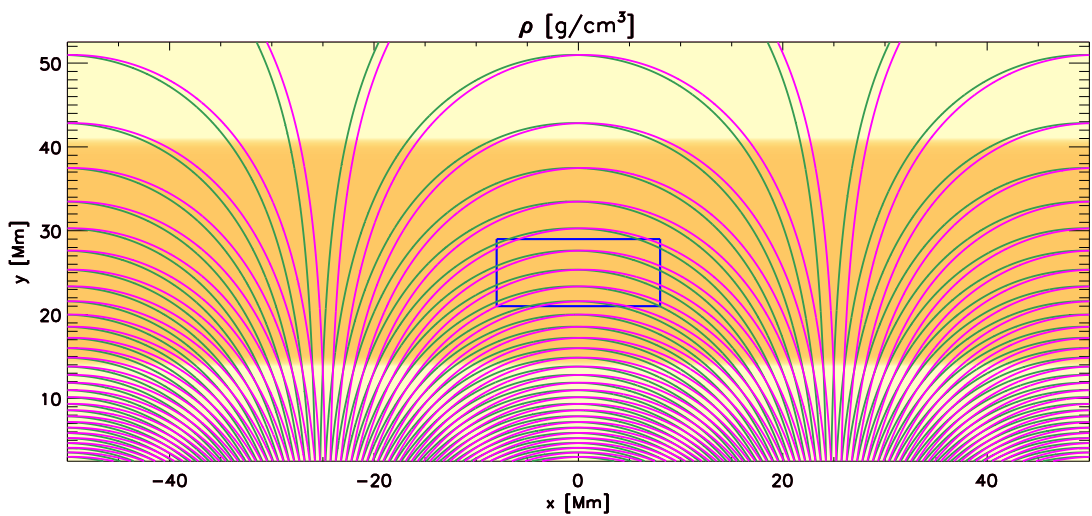}
\caption{Map of density in the z=0 plane in the initial condition of the MHD simulations. The maximum in this map is at $\rho=6\times10^{-14}$~$g/cm^{-3}$.
Green lines show the magnetic field lines for $z<0$ and magenta lines for $z>0$ when we set $x_c=0.5$~Mm.}
\label{initlarge}
\end{figure*}
Fig.~\ref{initlarge} shows a density map on the z=0 plane,
where we also plot magnetic field lines for $z>0$ (magenta lines) and for $z<0$ (green lines) when we set $x_c=0.5$~Mm.
The central blue box is the domain around the location  ($x_J$,$y_J$)
where we focus our investigation.
We setup this simulation in a grid of $4096\times2048\times64$ cubic cells where we have a spatial resolution of
$\Delta x=\Delta y=\Delta z=0.024$~Mm.
We use outflow boundary conditions at the $y-$boundaries and periodic boundary conditions at the $x$ and $z$ boundaries. The outflow $y-$boundaries are not force-free and thus the presence of the high density regions is key to maintain the centre of the domain unperturbed for a long enough time to study the evolution around ($x_J$,$y_J$). Moreover, we do not use a magnetic field divergence cleaning technique, as the magnetic field evolution is computed using the constrained transport approach \citep{Balsara_1999JCoPh.149..270B}.

\subsection{Numerical simulations}

\begin{figure*}
\centering
\includegraphics[scale=0.185]{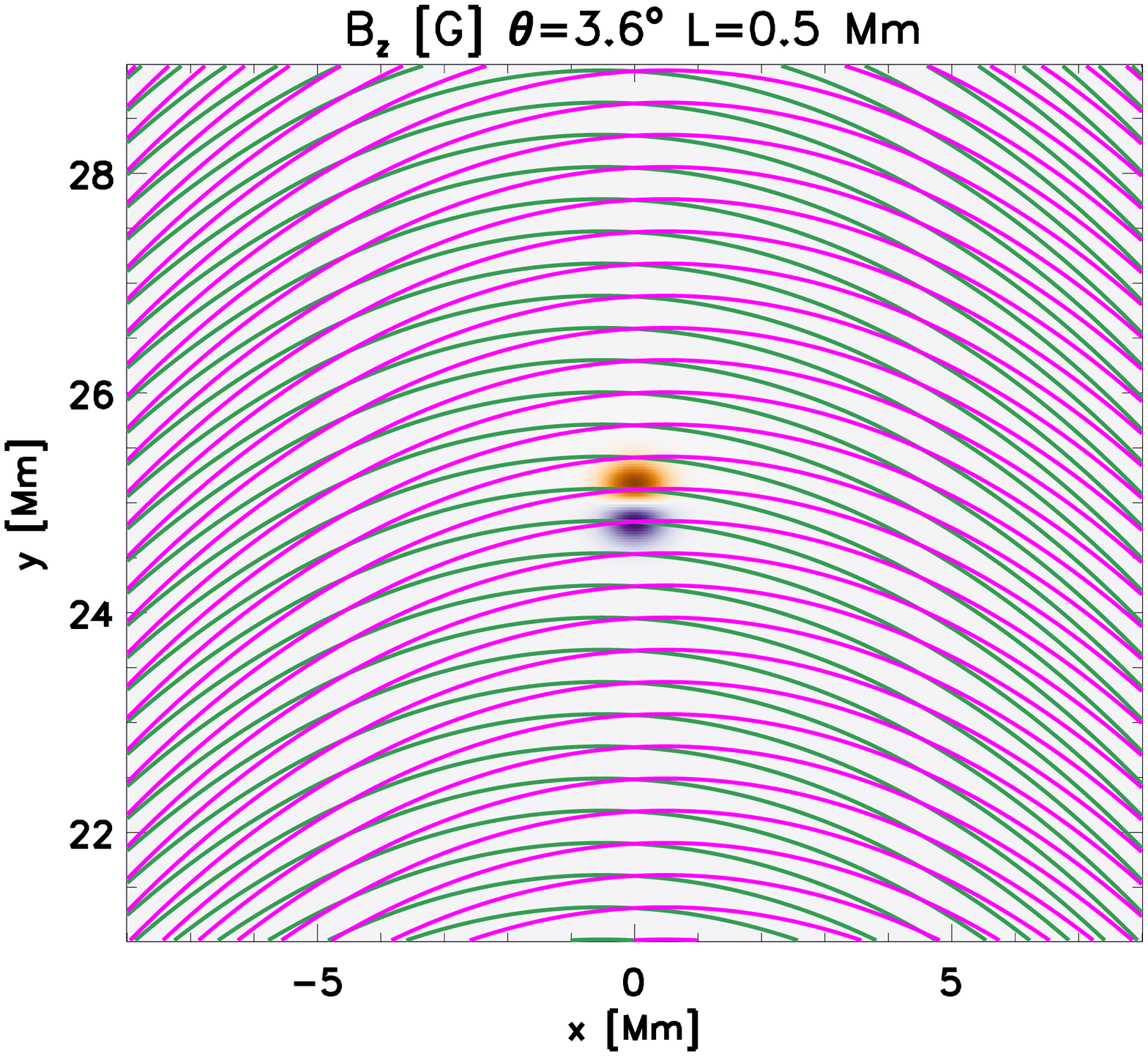}
\includegraphics[scale=0.185]{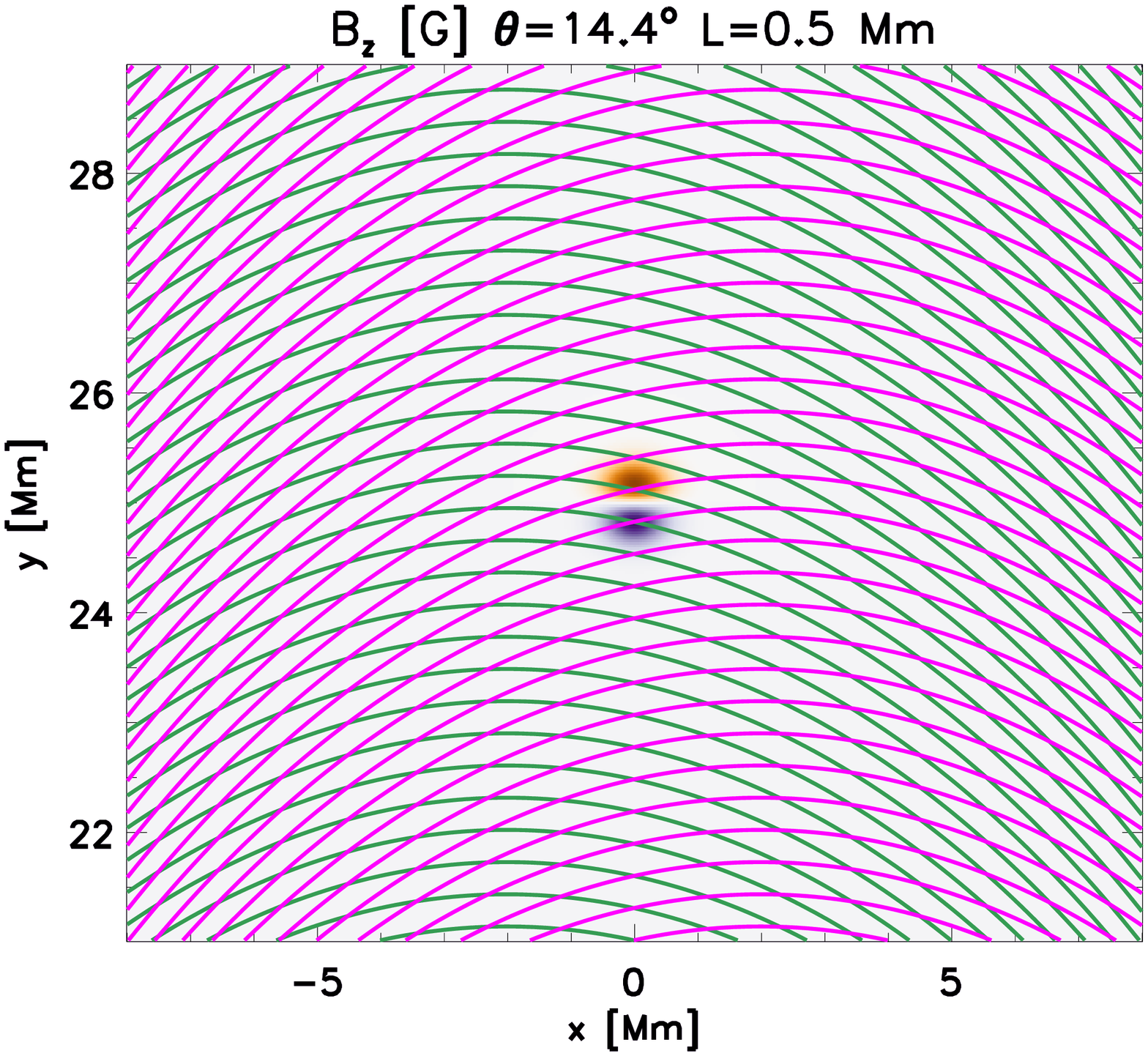}
\includegraphics[scale=0.185]{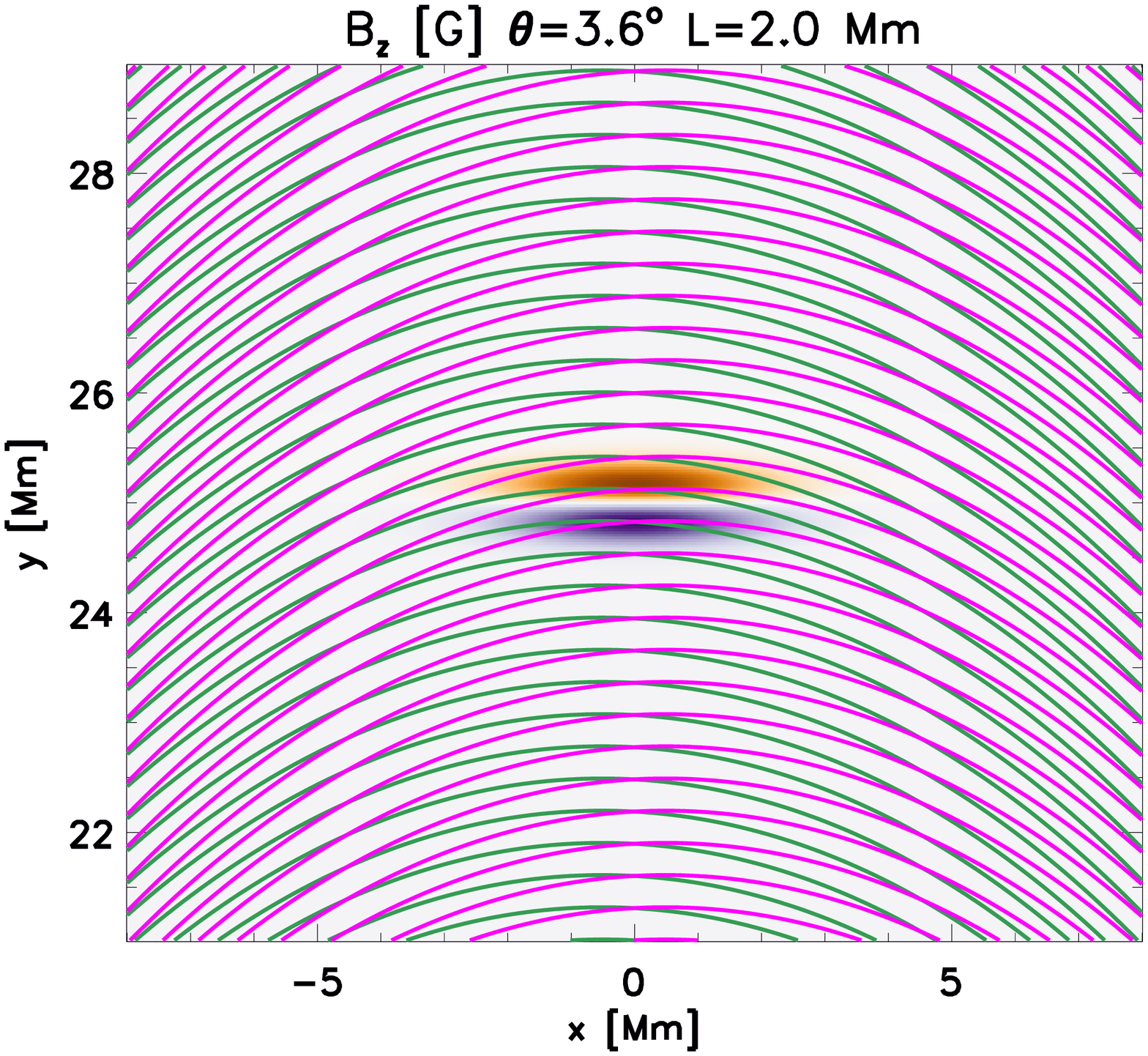}

\includegraphics[scale=0.205]{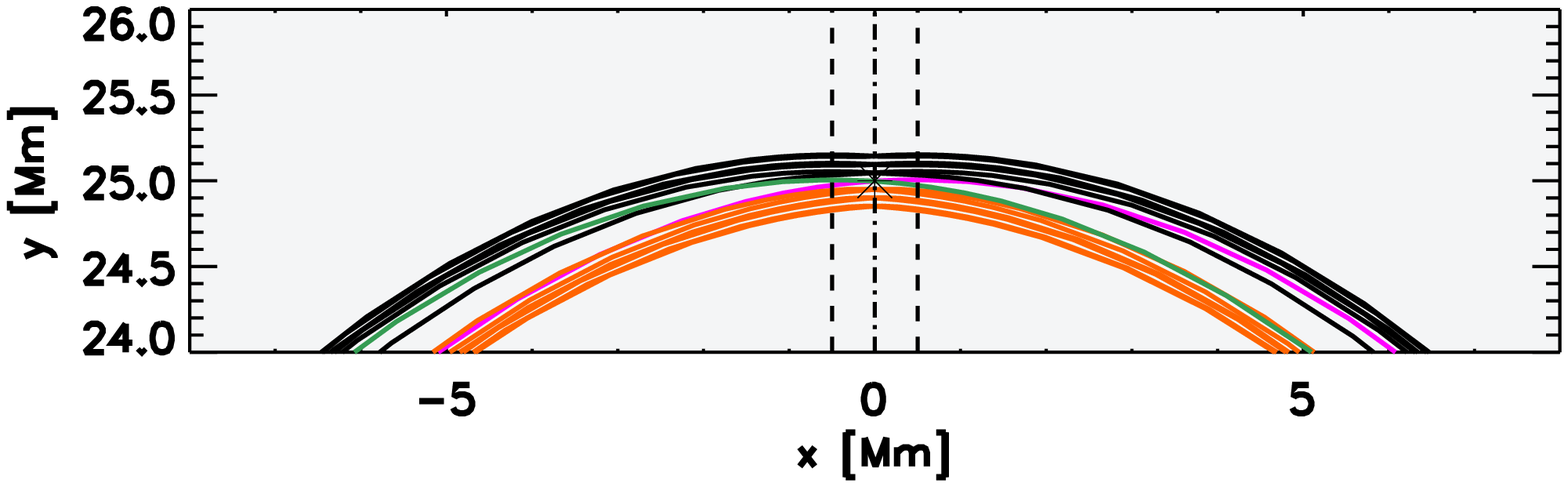}
\includegraphics[scale=0.205]{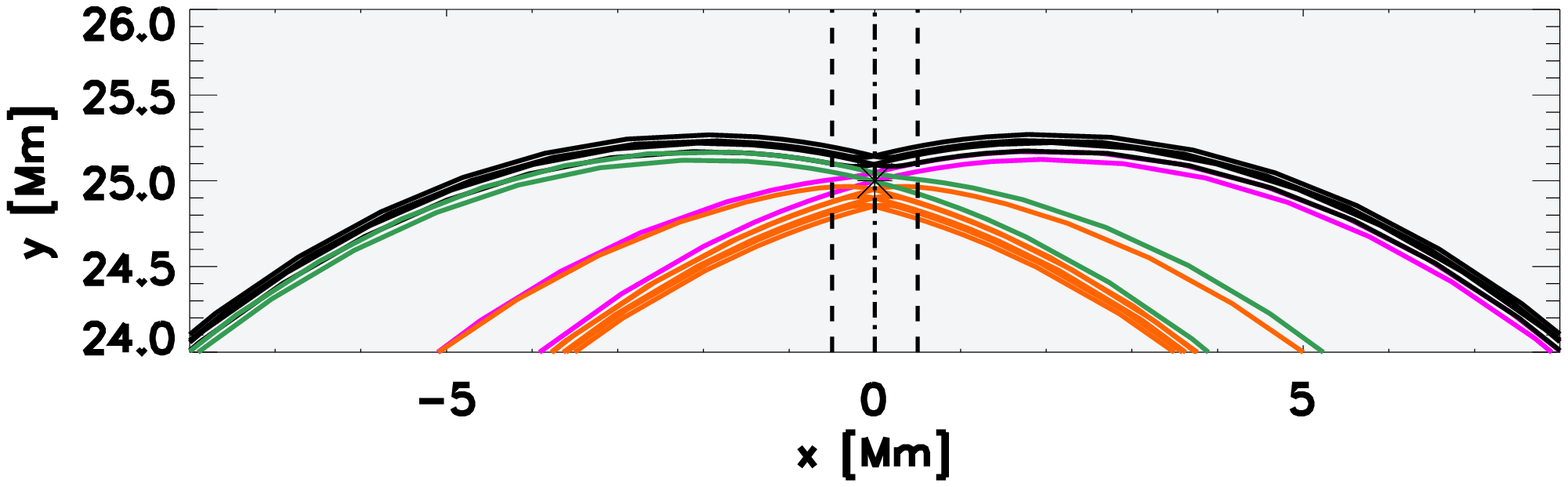}
\includegraphics[scale=0.205]{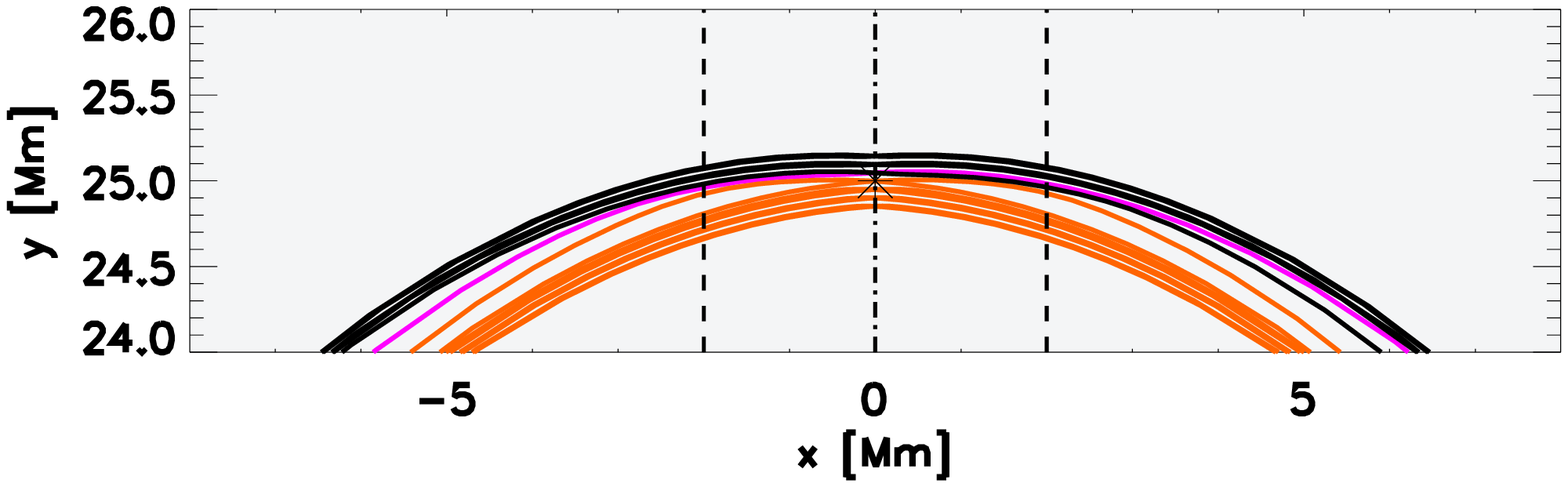}

\includegraphics[scale=0.205]{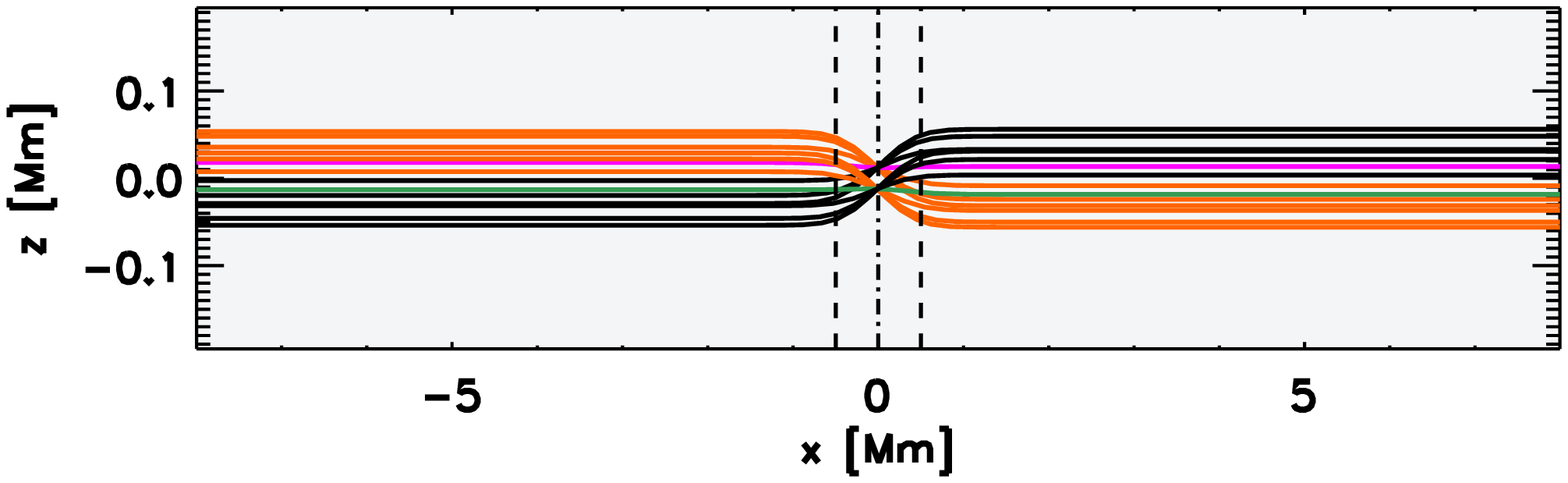}
\includegraphics[scale=0.205]{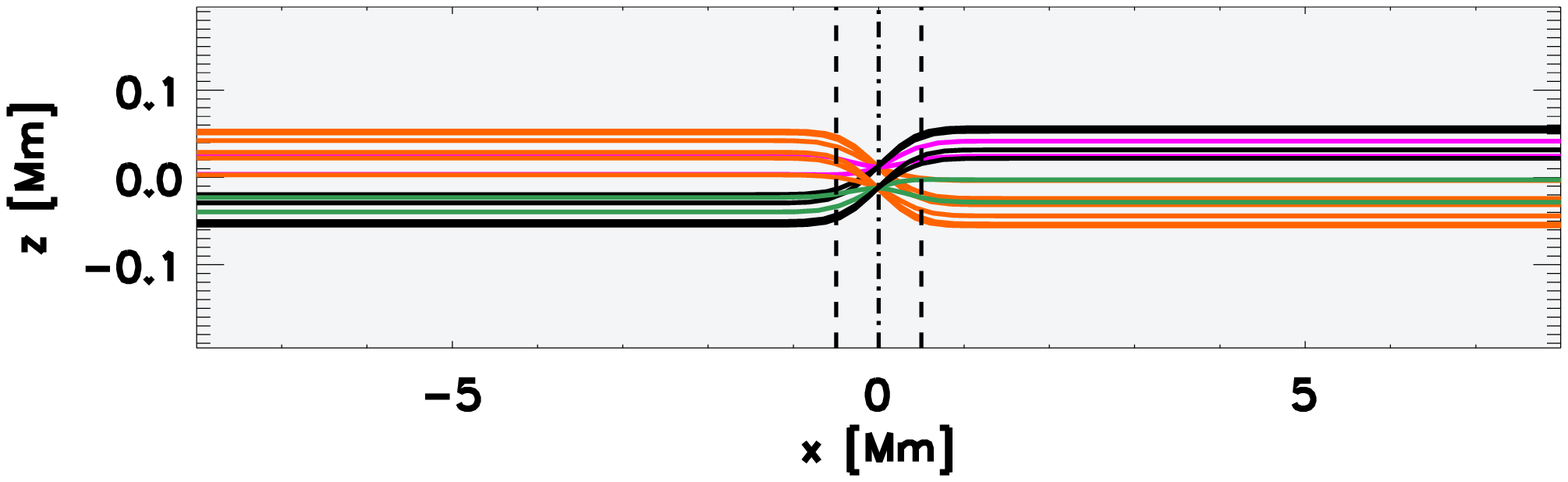}
\includegraphics[scale=0.205]{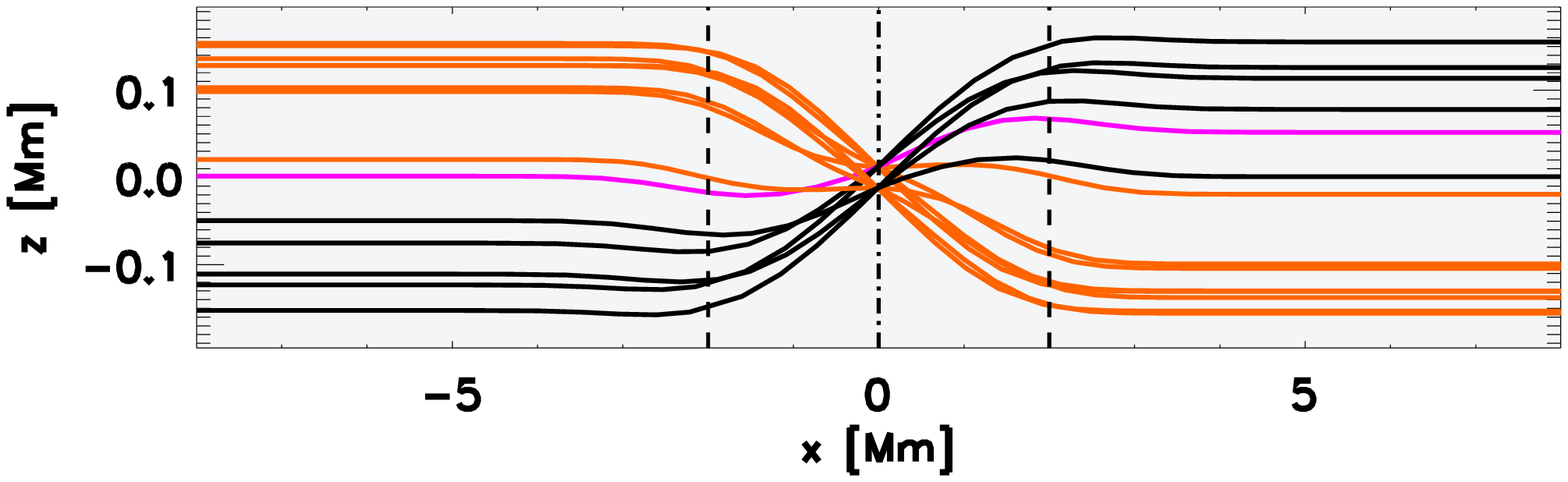}

\caption{The upper row shows maps of $B_z$ in the initial conditions of our simulations with superimposed magnetic field lines (green for $z<0$ and magenta for $z>0$), where we set
$x_c=0.5$~Mm and $L_{\parallel}=0.5$~Mm (left hand side column),
$x_c=2.0$~Mm and $L_{\parallel}=0.5$~Mm (central column),
$x_c=0.5$~Mm and $L_{\parallel}=2.0$~Mm (right hand side column) on the plane $z=0$.
In the two lower rows we show some representative magnetic field lines projected onto the x-y plane and x-z planes. Magnetic field lines are coloured black and orange if they cross the z=0 plane, and magenta and green if they do not.}
\label{simthl}
\end{figure*}

Fig.~\ref{simthl} shows some of the different initial configurations we consider. In the top row panels, we show maps of $B_z$ with over plotted some magnetic field lines. 
We take the LHS column simulation as the reference one. The central column simulation differs from the reference one for the angle between the magnetic field lines of the two flux systems (i.e. larger values of the parameter $x_c$) and the RHS column simulation for the wider region where a $B_z$ is present (i.e. larger values of the parameter $L_{\perp}$).
In the bottom panels we zoom into the region around $(x_J,y_J)$ and
we plot some magnetic field lines projected onto the $x-y$ and $x-z$ plane
and these are coloured black and orange if they cross the $z=0$ plane
(i.e. reconnected magnetic field lines) and are coloured magenta and green if they remain in the same arcade system.
It should be noted that in this configuration, $B_z$ bridges the two arcade system across the current sheet over a length of $\sim0.1$ $Mm$ which corresponds to 4 cells.
In this set of simulations we vary the parameters $x_c$ and $L_{\parallel}$.
Changing the parameter $x_c$ the two arcade systems become more shifted in $x$ and the angle $\theta$ between the field lines from different arcades increases. We consider three values of $x_c$ such as to have $\theta=3.6^{\circ}$, $\theta=7.2^{\circ}$, and $\theta=14.4^{\circ}$. As for the extension of the $B_z$ distribution, we run simulations with $L_{\parallel}=0.5$~Mm, $L_{\parallel}=1.0$~Mm, and $L_{\parallel}=2.0$~Mm.
The simulation with $\theta=3.6^{\circ}$ and $L_{\parallel}=0.5$~Mm is used as reference and any other simulation varies only one of the two parameters, leading to a total of 5 simulations.

\subsection{Evolution in the reference simulation}

\begin{figure}
\centering
\includegraphics[scale=0.40,clip,viewport=45 10 665 330]{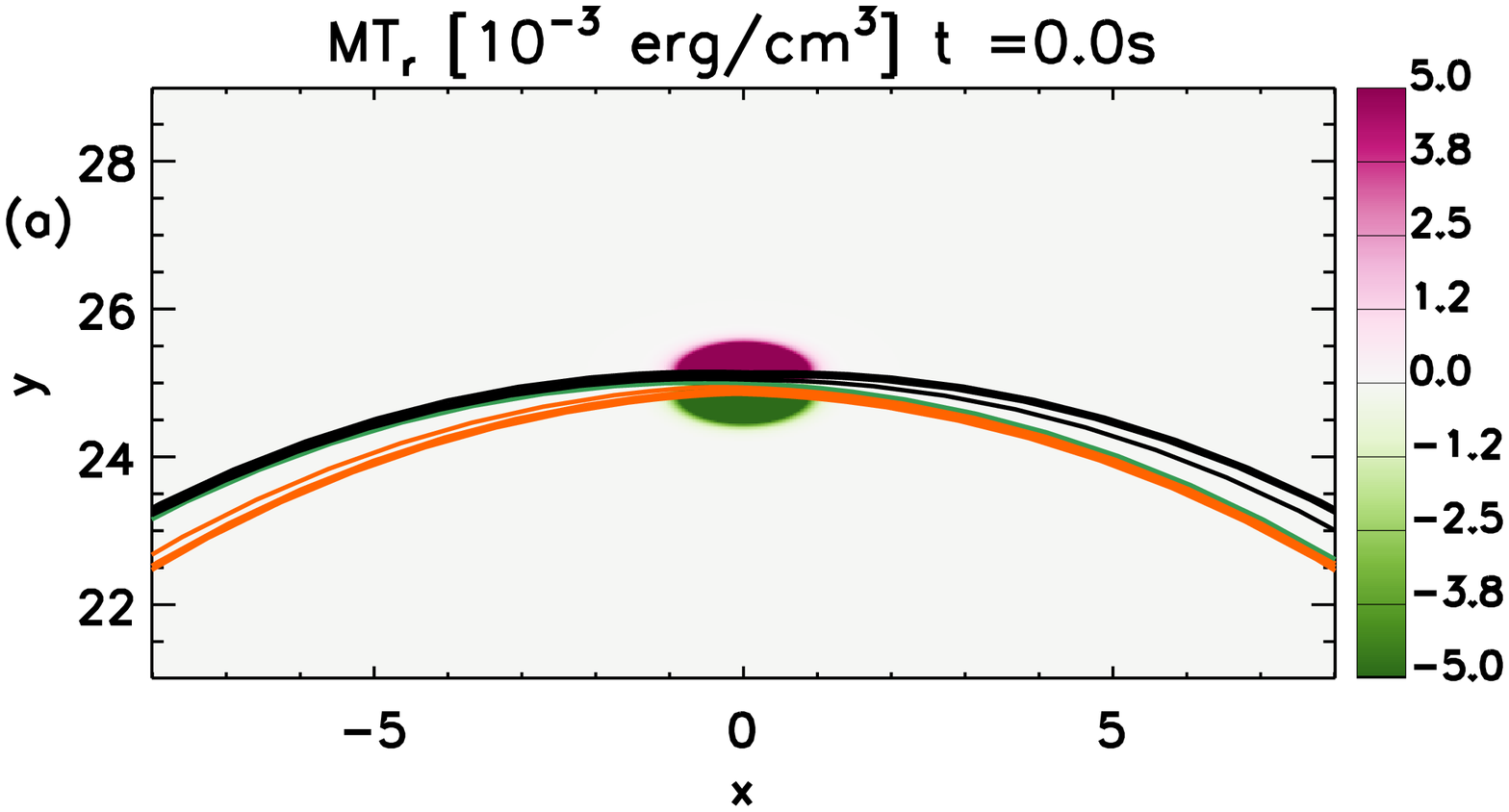}

\includegraphics[scale=0.40,clip,viewport=45 10 665 330]{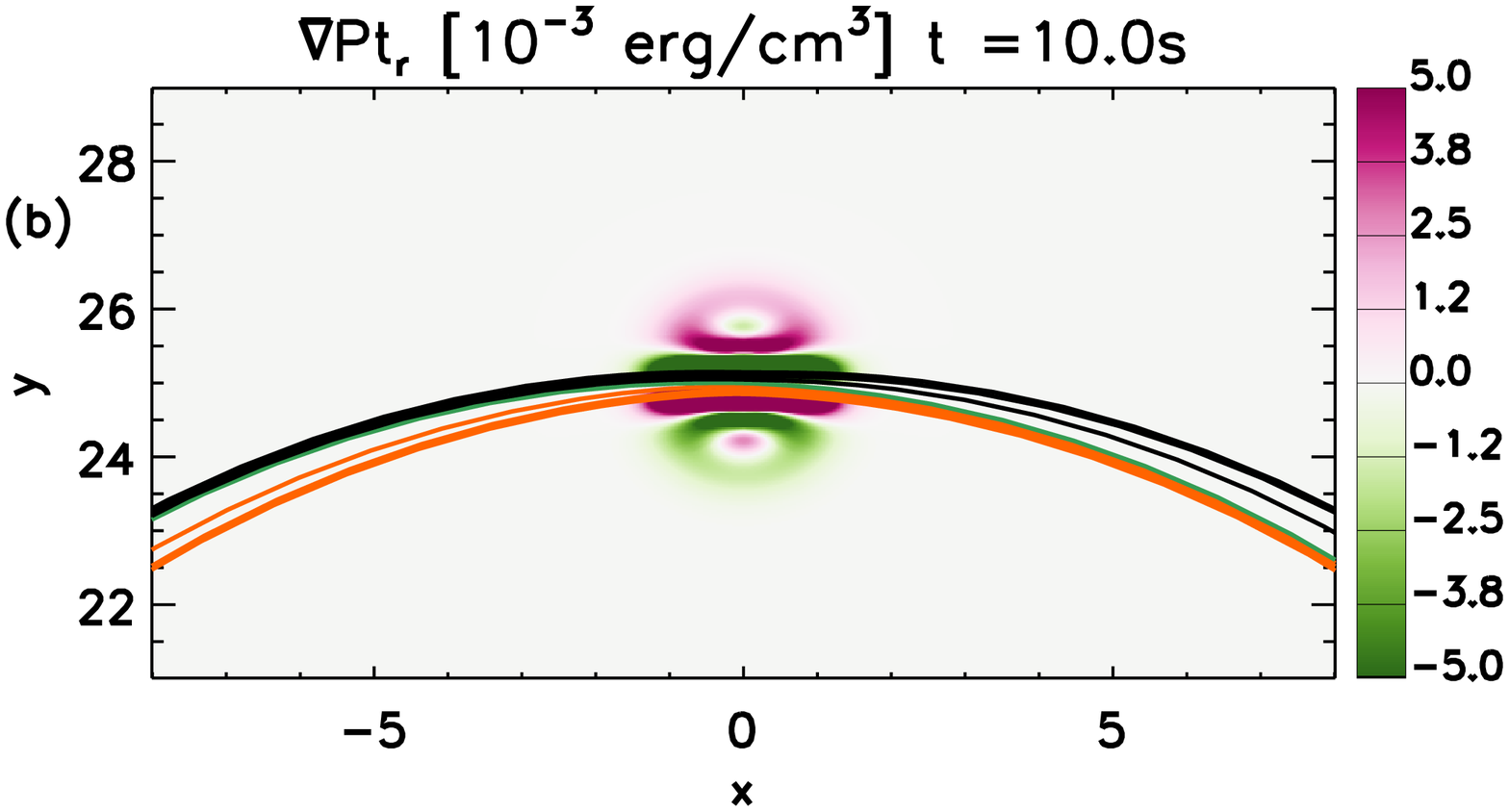}
\caption{(a) Map of the difference in the radial component of the magnetic tension between our simulation with $x_c=0.5$~Mm and $L_{\parallel}=0.5$~Mm and the analogous magnetic tension in a configuration where we set $B_{z0}=0$ on the plane $z=0$.
(b) Radial component of the gradient of the total pressure at $t=10$~$s$ in the same simulation as (a).
In both panels are superimposed some representative field lines.}
\label{forcejet}
\end{figure}

Our reference simulation has $\theta=3.6^{\circ}$ and $L_{\parallel}=0.5$~Mm. Its evolution is 
qualitatively similar to the other simulations and we thus illustrate only this one in greater detail.
Fig.~\ref{forcejet}a shows the radial component of the out of equilibrium magnetic tension in the initial condition. To measure this quantity, we compare this magnetic configuration with an analogous one where we have used $B_{z0}=0$.
We find that the magnetic tension pushes the plasma outwards from the reconnection region being positive above $y_J$ and negative below. The intensity of the magnetic tension above/below $y_J$ pushing the plasma upwards/downwards is roughly the same. 
This is an important difference in the construction of this MHD model with respect to the analytical construction sketched in Fig.~\ref{fig2}, where the magnetic tension inwardly directed is larger than the one outwardly directed. However, this difference holds only for the initial condition and as soon as the system starts evolving the inwardly directed magnetic tension becomes slightly, but consistently, larger than the one outwardly directed.

As soon as the plasma starts moving, a total pressure gradient develops.
Fig.~\ref{forcejet}b shows the radial component of the total pressure gradient at $t=10$~$s$. The total pressure distribution presents a complex structure, generally contrasting the magnetic tension and it thus acts to slow down the plasma flows, as it is half as intense as the magnetic tension.  

\begin{figure}
\centering
\includegraphics[scale=0.40,clip,viewport=45 10 665 330]{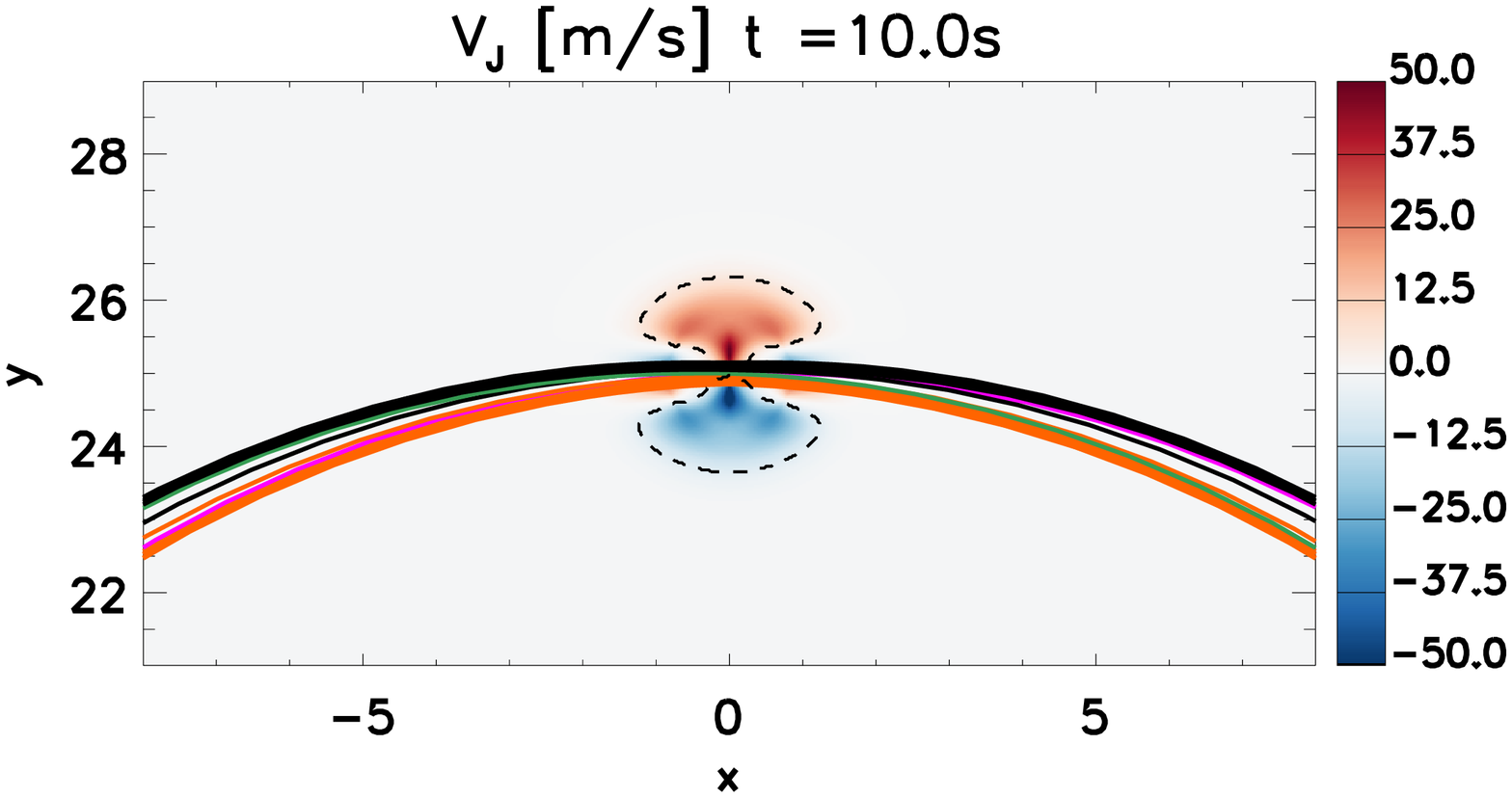}

\includegraphics[scale=0.40,clip,viewport=45 10 665 330]{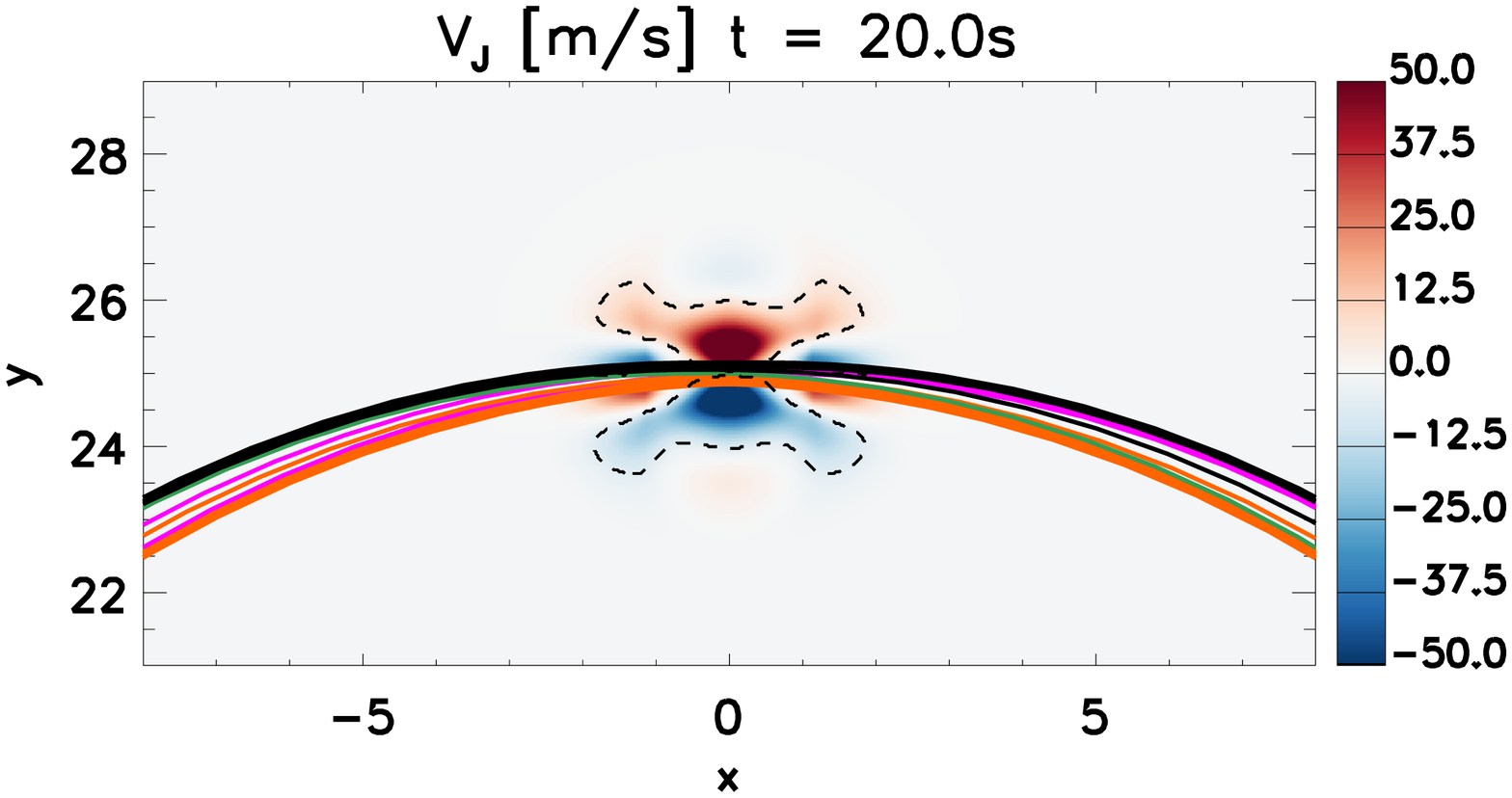}

\includegraphics[scale=0.40,clip,viewport=45 10 665 330]{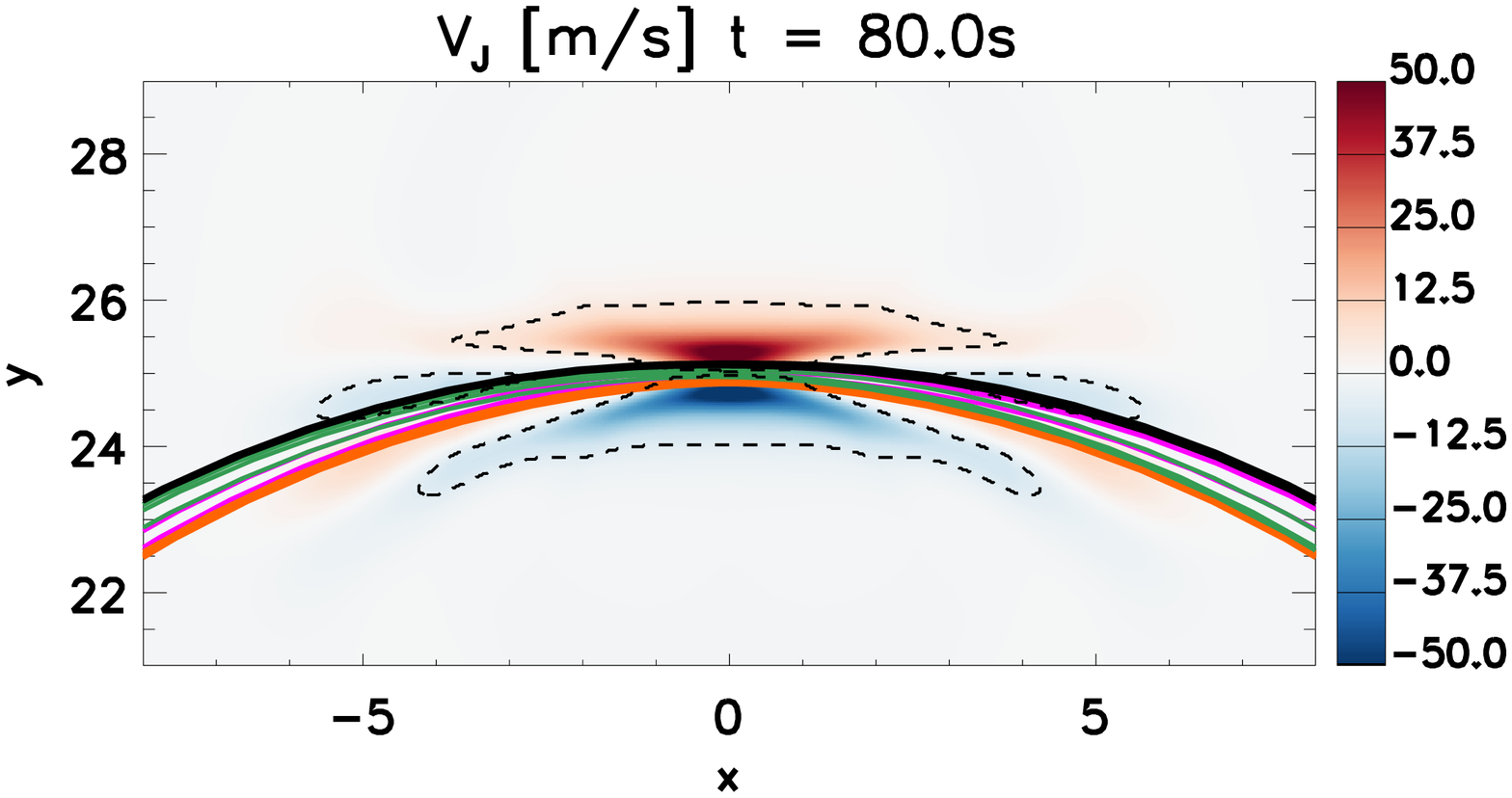}
\caption{Maps of the velocity of the jets,$V_J$, defined
as the projection of the radial velocity onto the direction perpendicular to the magnetic field
at three different times in the simulation with $x_c=0.5$~Mm and $L_{\parallel}=0.5$~Mm on the $z=0$ plane.
In all panels some representative field lines are superimposed.
Black dashed contours show the region where the jet speed is above $5$~$m/s$ and it is positive for the region above $y_J$ and negative for the region below $y_J$.}
\label{vrjet}
\end{figure}

Such forces lead to a bidirectional jet evolution.
One jet propagates outwards with respect to the curvature of the magnetic field (above $y_J$), while the other jet propagates inwards (below $y_J$).
The evolution of the simulation can be summarised in three key stages represented in Fig.~\ref{vrjet}, where we plot the projection of the radial velocity onto the direction perpendicular to the magnetic field, $V_J$, at different times.
The dashed contours in Fig.~\ref{vrjet} limit the region above and below the origin of the jets where $V_J$ is, respectively, positive or negative with a magnitude above $5$~$m/s$.
This is the component of the velocity 
that moves outwards from ($0$,$y_C$) that is caused by the magnetic tension.
In the very early stage of the evolution the quantity $V_J$ captures a magnetoacoustic perturbation moving away from the reconnection point and that is not relevant for our study. This perturbation is still visible at $t=10$ $s$ as two semicircles. However, at the same time, within $0.5$ $Mm$ from the ($0$,$y_C$) point we find higher velocity where the plasma moves outwards accelerated by magnetic tension. At a later time, $V_J$ (and its contour) properly describes the jets evolution, as their speeds then drop as they start interacting with the background medium.
Such velocities, modest in comparison to what normally observed in the solar corona, can be explained by the model parameters that have been chosen for these numerical experiments. In particular, the high plasma density, the plasma $\beta$ higher than solar active regions and the low magnetic field intensity contribute to forming jets which are slower than the typical velocities observed in the solar corona.
It should be noted, however, that during the evolution we find a decrease in magnetic energy in the proximity of the reconnection point that is fully converted in kinetic (jets) and  thermal (compression) energy. The magnetic energy conversion is approximately the amount of magnetic energy initially stored in the $B_z$ component of the magnetic field. These energetic considerations indicate that the energy of the jets depends on how much free magnetic energy is available in the initial condition.

In the last phase, the jets continue expanding mostly in the y direction for some time, and at $t=20$ $s$ we find two larger regions with positive and negative velocities.
The jets deceleration phase is longer than the acceleration phase, but still effective within the timescale of our simulation, as the jets' speed drop to half of their maximum speed in $75$~$s$.
In this last phase, represented by $t=80$ $s$ we find that the jets slow down and as a result, the motion also spreads along the x direction. During this evolution the magnetic field lines reconfigure in a way that the intensity of the magnetic tension decreases.

\begin{figure}
\centering
\includegraphics[scale=0.35]{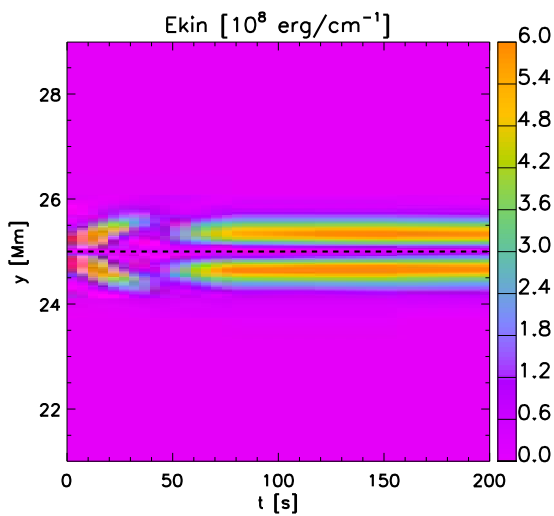}
\caption{Time-space map of the distribution of the integral of $E_J$ across the x-direction in the simulation with $x_c=0.5$~Mm and $L_{\parallel}=0.5$~Mm on the $z=0$ plane.}
\label{kineticener}
\end{figure}

In order to measure the asymmetry between these jets
we consider the kinetic energy associated to 
the jet velocity, i.e. $E_J=0.5\rho V_J^2$,
and we compute the integral across the horizontal cut at various y-coordinates in the region within the dashed contours in Fig.~\ref{vrjet}.
Fig.~\ref{kineticener}a shows a map of the quantity $E_J$ as a function of time and y-coordinates.
We find that the kinetic energy concentrates in the regions just below and above $y_J$ (dashed line in Fig.~\ref{kineticener}) and it decreases during the MHD simulation because of the jet deceleration.
The initial structures ($t<15$~$s$) propagating symmetrically from $y_J$ are due to the fast magnetoacoustic perturbations
and already at $t=15$ $s$ are far enough from the origin of the jets.
After $t=15$ $s$, instead, the energy directly associated with the jets becomes predominant in the map, as two bands that grow and then remain approximately at the same y location from $t=50$ $s$.
In particular, the lower region consistently shows higher kinetic energy values.

\begin{figure}
\centering
\includegraphics[scale=0.35]{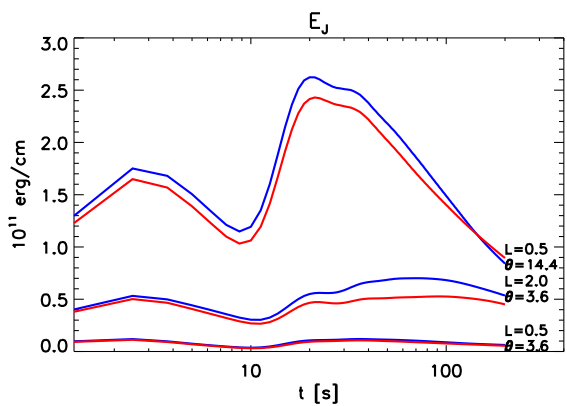}
\caption{Time evolution of the integral of $E_J$ in the region above (red lines) and below (blue line) $y_J$ for 3 different simulations, whose parameters are listed in the plot.}
\label{ejtime}
\end{figure}

Fig.~\ref{ejtime} shows the evolution of the quantity $E_J$ integrated in the region above the $y_C$ (red lines) and below it (blue lines) for three different simulations.
We include for reference the other two simulations 
where we change either the angle $\theta$ or the extent of the $B_z$ distribution $L_{\parallel}$.
For all the simulations, the evolution always remains asymmetric, where the inward propagating jets always carry more energy than the outward counterparts.

During the very first stage of the evolution, when the magnetoacoustic perturbations are still present, the evolution is still symmetric, but it becomes increasingly more asymmetric as the jets accelerate ($t<20$~$s$).
Additionally, we find that the energy of the jets is different for different configurations as it increases when we either increase $L_{\parallel}$ or $\theta$. In particular the simulation with higher tilt angle between the two arcade systems lead to the most energetic jets.
Naturally, these stronger jets incur in the most effective damping as they interact with the background medium.
Moreover, as the magnetic field intensity is higher below the reconnection point, the inwardly directed jets are more effectively decelerated. This effect becomes more important the larger the displacement of the jets.
For this reason the inward jet of the simulation with $\theta=14.4^{\circ}$
eventually gets slower than the outward jet  ($t\sim150$~$s$).

It should be noted that the initial condition of this MHD model shows a nearly symmetric configuration of the initial magnetic tension forces that generate the jets. However, the asymmetry in the jets still develops and increases in times as they accelerate. This corroborates that the curvature of the magnetic field affects the symmetry of the jets not only when they are generated, but also during their acceleration phase.

\subsection{Dependence on model parameters}

In order to inspect the model we have described in Sect.\ref{analytical}, we hereby look into the dependence on the $\theta$ angle between the reconnecting field lines (tilt angle) and the dependence on the width of the $B_z$ distribution, i.e. $L_{\parallel}$.
We therefore consider the simulations where we increase either parameters.
In order to compare how the asymmetry of the jets develops, for all these simulations we measure the ratio of the integral of $E_J$ below and above the reconnection point as a function of time (see Fig.~\ref{ejtime}).

\begin{figure}
\centering
\includegraphics[scale=0.35]{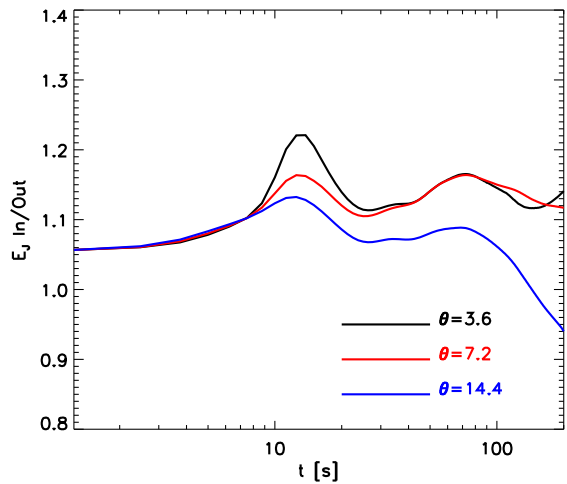}
\includegraphics[scale=0.35]{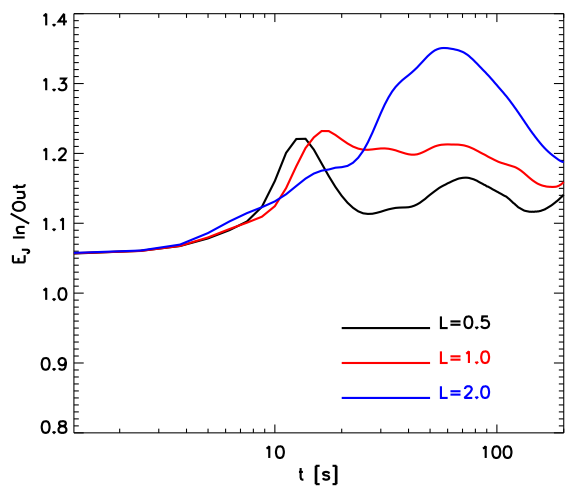}
\caption{(a) Time evolution of the asymmetry of the jets (ratio between the integral of $E_J$ below and above $y_J$) for the three simulations differing for the angle $\theta$.
(b) Time evolution of the asymmetry of the jets for the three simulations differing for the width of the intial $B_z$ distribution  $L_{\parallel}$.}
\label{thetapar}
\end{figure}
Fig.~\ref{thetapar}a shows how this ratio evolves for
the simulations where we vary the $\theta$ parameter.
In all cases, the evolution shows a similar pattern in which the ratio initially increases, it then reaches a maximum, and it finally decreases during the jets deceleration phase.
In this setup, the peak of asymmetry is dependent on the angle, as predicted in Sect.\ref{analytical}. Indeed, the simulations with higher angle consistently show a lower degree of asymmetry at all times.

Fig.~\ref{thetapar}b shows the evolution of 
the asymmetry for the simulations where we vary the width of the $B_z$ distribution. Also in this case, the evolution
shows a similar pattern of an initial increase followed by
a maximum, and finally a decrease.
However, this happens on different time scales,
where the simulation with $L_{\parallel}=2$ is the slowest in reaching a maximum.
The comparison with our results in Sect.~\ref{analytical} is here less evident,
as the width of the $B_z$ distribution is not obviously associated with the retracting length.
However, for both parameters the larger is their value the larger portion of the curved magnetic field lines is involved and the more the global curvature of the loop becomes important.
At the same time, in Sect.~\ref{analytical} we assumed that the retracting length varies in time as the reconnection occurs locally but the reconfiguration expands from there. Such behaviour is not reproduced in these MHD simulations where the parameter $L_{\parallel}$ only affects the initial condition.

However, none of these simulations reproduces the regime described in Sect.\ref{analytical} where there is no outward directed magnetic tension force. This is because the reconnected magnetic field is inherent to the initial condition and we thus always have an outward directed magnetic tension.
Another significant difference between this approach and the one in Sect.\ref{analytical} is that in these simulations the width of the region affected by the reconnection is determined from the beginning and it does not grow in time as in Sect.\ref{timeevolanalytical}.
Because of this the asymmetric growth of the magnetic tension force described in Sect.\ref{timeevolanalytical} cannot be reproduced by the MHD simulations, where, in contrast, the asymmetry in the magnetic tension force is maximum at the initial condition and it then slightly decreases in time.

\section{Discussion and conclusions}
\label{conclusions}

In this work we have investigated the asymmetric nature of reconnection jets from curved magnetic field lines. The nanojets in the solar corona \citep{AntolinNature} are an example of these phenomena.
We focused on some probable explanations for such asymmetry and we have addressed this problem from two points of view, first using a simple geometrical model to explain why we expect asymmetric jets, and second, using MHD simulations to analyse the evolution of these phenomena after the reconnection took place.

The observations show that nanojets show a preferential direction of propagation inward with respect to the curvature of the hosting coronal loop. Our investigation suggests that such dynamics can be explained by the larger inward magnetic tension forces, generated when two curved magnetic field lines reconnect. 

Using our simple geometrical model, we have shown than the global curvature of the loop contributes to the local magnetic field line curvature that is formed as a consequence of the magnetic reconnection. Therefore the curved magnetic field environment breaks the symmetry for the otherwise perfectly bidirectional nanojets, and causes the inwardly directed flow to exceed the outwardly directed one.

These asymmetric effects become less relevant as the misalignment angle between the reconnecting magnetic field line increases, since this leads to higher local curvatures at the reconnection point that exceed the global curvature of the coronal loop.
This effect, however, can be compensated when the retracting length of the magnetic field lines involve a larger region, as in this case again, the global curvature dominates over the local one.

In our geometrical model we also find two different possible regimes, one where the nanojets are generated by asymmetric tension forces, and the second one where the outward tension simply cannot be generated.
Our analytical model is mostly based on the local curvature of the field lines at both sides of the reconnection point, which is here controlled by the global curvature of the field lines. Hence our analysis is not only valid for loops curvature but can easily be generalised to any kind of factor leading to local curvature, such as loop braiding. Along this line of thought, if the local curvature introduced by braiding is more important than that set by loop curvature, it is likely that nanojets are not only inwardly directed, but may have any direction while still being unidirectional.

When we approach the problem from an MHD perspective we find consistent results, but also interesting differences.
In order to study the jets in MHD we devise a two-arcades system that are shifted one near the other separated by a current sheet.
A connecting magnetic field is enforced to model the post-reconnection magnetic configuration and the magnetic tension generated in this way pushes the plasma away leading to two jets, one inwardly and one outwardly directed.
In the MHD framework, we recover some of the analytical result, such as that the jets become more symmetric as the misalignment angle between the magnetic field lines increases, and that they become more asymmetric with a wider regions across which magnetic field lines are connected.
Crucially, the MHD simulations confirm that the inward jets are consistently more energetic than the outward ones.

On the other hand, MHD simulations cannot reproduce the regime where no outward flow exists and this is because the imposed magnetic field inherently causes a bidirectional jet.
Additionally, the jets in the MHD simulations show a significantly smaller asymmetry, as the inward jets are not orders of magnitude more energetic, as prescribed by the analytical model. 
This is probably because the MHD model starts with a symmetric force that triggers the jets and it does not comprehensively describe the time evolution of the magnetic reconnection.

In the follow up of this work, we aim at running MHD simulations with an explicit resistivity term that allows for studying simultaneously in a full MHD framework the magnetic reconnection and the evolution of the jets. We expect such numerical experiments to develop more complex flows, as the magnetic field diffusion and the temporary cancellation of the magnetic field before the reconnection will lead to plasma flows along the magnetic field lines because of the magnetic pressure gradient. Additionally, the magnetic diffusion inevitably leads to ohmic heating which will trigger further plasma flows.

In this work we find that when the reconnection angle is small enough and the region involved in the reconnection is large enough, nanojets can be substantially asymmetric and this reconciles this model with the observational evidences found in \citet{AntolinNature}.
However, in \citet{AntolinNature} the large majority of the observed jets were unidirectional and such high degree of asymmetric is not matched in the MHD simulations here presented.

In conclusion, this modelling work establishes as proof of concept that the curvature of magnetic structures affects the symmetry of local reconnection jets, but a more complete MHD model needs to be developed to explain fully understand this mechanism and bridge the gap with observations.
Also, additional observations of nanojets are needed to consolidate this description.
In particular, high resolution observations and accurate reconstructions of the coronal magnetic field are necessary to ultimately validate this approach.
\begin{acknowledgements}

P.A. acknowledges funding from STFC Ernest Rutherford Fellowship (No. ST/R004285/2).
This work used the DiRAC@Durham facility managed by the Institute for Computational Cosmology on behalf of the STFC DiRAC HPC Facility (www.dirac.ac.uk). The equipment was funded by BEIS capital funding via STFC capital grants ST/P002293/1, ST/R002371/1 and ST/S002502/1, Durham University and STFC operations grant ST/R000832/1. DiRAC is part of the National e-Infrastructure.
PLUTO was developed at the Turin Astronomical Observatory in collaboration with the Department of Physics of the Turin University.

\end{acknowledgements}

\bibliographystyle{aa}
\bibliography{ref}

\end{document}